# GRAVITATIONAL INSTABILITY IN COLLISIONLESS COSMOLOGICAL PANCAKES [1]

Azita Valinia,[2,3,4] Paul R. Shapiro,[5] Hugo Martel,[5] and Ethan T. Vishniac[5]

## ABSTRACT

The gravitational instability of cosmological pancakes composed of collisionless dark matter in an Einstein-de Sitter universe is investigated numerically to demonstrate that pancakes are unstable with respect to fragmentation and the formation of filaments. A "pancake" is defined here as the nonlinear outcome of the growth of a 1D, sinusoidal, plane-wave, adiabatic density perturbation. We have used high resolution, 2D, N-body simulations by the Particle-Mesh (PM) method to study the response of pancakes to perturbation by either symmetric (density) or antisymmetric (bending or rippling) modes, with corresponding wavevectors $\mathbf{k}_s$ and $\mathbf{k}_a$ transverse to the wavevector $\mathbf{k}_p$ of the unperturbed pancake plane-wave. We consider dark matter which is initially "cold" (i.e. with no random thermal velocity in the initial conditions). We also investigate the effect of a finite, random, isotropic, initial velocity dispersion (i.e. initial thermal velocity) on the fate of pancake collapse and instability.

Our results include the following: (1) For "cold" initial conditions, pancakes are gravitationally unstable with respect to all perturbations of wavenumber $k \gtrsim 1$ (where $k = \lambda_p/\lambda$, and $\lambda_p$ and $\lambda$ are the wavelengths of the unperturbed pancake and of the perturbation, respectively). This is contrary to the expectations of an approximate, thin-sheet energy argument applied to the results of 1D pancake simulations. The latter predicts that unstable wavenumbers are restricted to the range $k_{\min} < k < k_{\max}$, where perturbations with $k < k_{\min} \sim 1$ are stabilized by Hubble expansion, while those with $k > k_{\mathrm{max}} > 1$ are stabilized by the 1D velocity dispersion of the collisionless particles along the direction of pancake collapse, within the region of shell-crossing. (2) Shortly after the pancake first reaches a nonlinear state of collapse, the dimensionless growth rate of the perturbation of pancake surface density by unstable transverse modes rises rapidly from the value of 2/3 for a linear density fluctuation in the absence of the primary pancake and levels off. This signals the onset of a new, *linear* instability of the *nonlinear* pancake, which grows as a power-law in time. (3) The index of this power-law time-dependence scales as $k^n$, where $n \approx 0.2 - 0.25$ for both the symmetric and antisymmetric modes. (4) The power spectrum of the perturbation in pancake surface density is strongly peaked during the linear phase of growth of unstable modes, at wavenumber $k = k_s$ for symmetric modes and $k = 2k_a$ for antisymmetric modes. (5) Eventually, the onset of nonlinearity is signaled by a decline of the growth rate and the production of clumps of large overdensity relative to that of the unperturbed pancake, located in or near the plane of the unperturbed pancake. In 2D, these "clumps" correspond to filaments, one per $\lambda_s$ (two per $\lambda_a$) for symmetric (antisymmetric) modes. (6) For equal initial amplitudes, the antisymmetric mode reaches nonlinearity later than the symmetric mode. (7) These filaments have azimuthally-averaged density profiles $\rho \sim r^{-m}$, $m = 1.1 \pm 0.1$. (8) Contrary to the expectations of the thin-sheet energy argument, pancakes in a collisionless gas of finite temperature are also gravitationally unstable with respect to all perturbations of wavenumber $k \gtrsim 1$. Finite temperature does not stabilize the pancake against perturbations of large wavenumber as predicted by the energy argument. Finite temperature *can* cause perturbations to decay prior to the collapse of the primary pancake, but once pancake caustics form, transverse perturbations grow at the same linear growth rate as for cold initial conditions.

*Subject headings:* cosmology: theory — dark matter — galaxies: formation — instabilities — large-scale structure of Universe

# 1. INTRODUCTION

Gravitational instability is believed to be the dominant process through which large-scale structure forms in the expanding universe. Observations reveal that galaxies and clusters of galaxies are concentrated in space within sheetlike and filamentary structures (e.g. de Lapparent, Geller, & Huchra 1988; da Costa et al. 1988). This observed structure is commonly assumed to result from the gravitational instability of initially small-amplitude, Gaussian-random-noise, primordial density fluctuations in an otherwise uniform medium. Depending on the shape of the initial power spectrum of these fluctuations and the type of matter which is assumed to dominate the total density at late times, the formation of structure may proceed in either a "top-down" or "bottom-up" fashion. In the "top-down" scenario, power at small wavelengths is absent at late times before nonlinear structure forms. As a result, large-scale structure forms first and fragments into smaller objects. In the "bottom-up" cosmogony, on the other hand, power is greater at small wavelengths than large, so small objects appear first and merge to form larger structures. The "top-down" scenario takes place, for example, in a universe dominated by Hot Dark Matter (HDM), whereas a "bottom-up" scenario results from a universe dominated by Cold Dark Matter (CDM). In a universe containing both types of dark matter (CHDM), a combination of these processes are at work to create the structure.

Numerical simulations provide a cosmic laboratory in which to test and understand the process of large-scale structure formation. Many numerical simulations focus on the evolution of a representative piece of the universe within the framework of a particular scenario, with the goal of comparing the structure that forms in the simulation with observations. The success of each model then depends on how well the simulated scenario produces the observed structure in the universe, in a statistical sense. However, the above approach is somewhat difficult to interpret in order to gain insight into the underlying nonlinear gravitational collapse process. A fuller understanding of the transition from linear to nonlinear phase can be achieved through idealized numerical simulations which provide insight to the behavior of gravitational instability in more realistic simulations.

A common feature of all cosmological scenarios involving Gaussian-random-noise fluctuations is the presence of flattened objects, or the so-called cosmological "pancakes." Pancakes were first shown by Zel'dovich (1970) to result from gravitational collapse of primordial adiabatic density perturbations in the context of a baryon-dominated universe. Pancakes are the first structures to form in cosmological models such as HDM, for which the power spectrum is truncated at some minimum wavelength $\lambda_{\min} \equiv 2\pi/k_{\max}$, where $k_{\max}$ is the maximum wavenumber. In such models, the smallest primordial nonlinear structures that form are pancakes and filaments of size $\lambda \sim \lambda_{\min}$, connected in a cellular network in which the cell walls are pancakes. Structures on scales $\lambda < \lambda_{\min}$ cannot be primordial, since there is no power at that scale in the initial conditions. Such structures can only originate from the transfer of power from scales larger than $\lambda_{\min}$ to scales smaller than $\lambda_{\min}$ by nonlinear mode coupling. Since this effect is negligible until the modes involved have reached nonlinearity, the pancakes must form first, before second order effects leading to their fragmentation become important. This is the best known example of a top-down scenario for structure formation. (For a review of the early work on pancakes by Zel'dovich and his collaborators, see Shandarin, Doroshkevich, & Zel'dovich 1983, and references therein.)

A question of great ongoing interest is whether pancakes also form and play an important role in cosmological models for which the power spectrum has no sharp cutoff at small wavelength. Indeed, such structures have been shown to form in numerical simulations of large scale structure formation which have widely different initial power spectra. Beacom et al. (1991), Melott & Shandarin (1989), and Kofman et al. (1992) have performed very high resolution, 2D simulations that show that pancakes and filaments form in models with initial density fluctuations which have a power law power spectrum $P(k) = |\delta_{\mathbf{k}}|^2 \propto k^n$ if the index $n \leq 0$ (i.e. $n \leq -1$ in 3D). Simulations in 3D with more realistic power spectra (such as that of CDM) also form pancakes and filaments (e.g. Cen & Ostriker 1992; Kang et al. 1994), although in this case identification of these structures is more difficult because such 3D simulations generally have much less length resolution than their 2D counterparts.

It is possible to understand these results qualitatively, in part, as follows. For the sake of discussion, we shall restrict our attention to models for which the power spectrum $|\delta_{\mathbf{k}}|^2$ in the matter-dominated era and in the linear regime decreases with increasing $k$, but slower than $k^{-3}$. A good example is the CDM power spectrum (see, for instance, Kolb & Turner 1990; Peebles 1993, and references therein), which varies approximately like $k^{-2}$ over a wide range of cosmologically interesting scales (from hundreds of kiloparsecs to tens of Megaparsecs). For such models, the contribution to the rms density fluctuation averaged over space from fluctuations of wavelength $\lambda = 2\pi/k$ is given by

$$\langle \delta_\lambda^2 \rangle \propto k^3 |\delta_{\mathbf{k}}|^2 \tag{1}$$

(Peebles 1980, eq. [26.5]). Equation (1) implies that these models have more *total* power at small wavelengths, though the power *per mode* is less. In this scenario, there is a fundamental length scale $\lambda_{NL}$ which is the comoving scale on which rms density fluctuations are just becoming nonlinear at the particular epoch we wish to study. In the absence of modes with wavelength $\lambda < \lambda_{NL}$, modes with $\lambda \gtrsim \lambda_{NL}$ would result in pancake formation, as, for example, in the HDM scenario. The presence of modes of wavelength $\lambda < \lambda_{NL}$ results in the formation of nonlinear structure on scales $\lambda < \lambda_{NL}$, however, since the rms density fluctuations are larger for smaller wavelength. Nevertheless, the remarkable fact is that the presence of this nonlinearity on scales $\lambda < \lambda_{NL}$ does not overwhelm the tendency to form pancakes on scales $\lambda \sim \lambda_{NL}$. This results from the fact that the peculiar velocity field on scales $\lambda_{NL}$ is dominated by wavelengths $\lambda > \lambda_{NL}$ (i.e. the velocity perturbation $\delta_{v,\mathbf{k}}$ associated with density fluctuation $\delta_{\mathbf{k}}$ is such that $\delta_{v,\mathbf{k}} \propto k^{-1}\delta_{\mathbf{k}}$). Hence, the assemblage of nonlinear structures of scale $\lambda < \lambda_{NL}$ into larger structures of scale $\lambda \sim \lambda_{NL}$ occurs in the presence of a large-scale peculiar velocity fluctuation due to $\lambda \gtrsim \lambda_{NL}$, which sweeps matter coherently into the same pancakes as would have been identified in a model without power on wavelength $\lambda < \lambda_{NL}$. The power on small scales, therefore, only affects the internal structures of the pancakes and filaments formed by the long wavelength modes.

The length scale $\lambda_{NL}$, therefore, plays essentially the same role as $\lambda_{\min}$ does for models like HDM. Hence, we expect the large-scale structure even in models like CDM to be made of extended structures (either pancakes or filaments) of size $\lambda \sim \lambda_{NL}$ connected in a network. Furthermore, analytical studies (Vishniac 1983a) and numerical simulations (Shandarin & Melott 1990) show that the transfer of power from modes with $\lambda < \lambda_{NL}$ to modes with $\lambda > \lambda_{NL}$ is very inefficient. Hence, the overall large-scale structures in these models should not depend on the presence of the small scale modes. This has been confirmed by numerical simulations (Little, Weinberg, & Park 1991; Martel, Shapiro, & Peebles 1996). The morphology of the large-scale structure formed in these simulations varies as modes with smaller wavelengths are progressively added to the initial conditions, until we reach the scale $\lambda_{NL}$, after which adding new modes does not change the outcome of the simulations. The three-dimensional PM N-body simulations of Little et al. (1991) reveal a network of connected filaments and pancakes. However, their numerical resolution is not sufficient to resolve the smallest pancakes of size $\lambda \sim \lambda_{NL}$. The simulations of Martel et al. (1996), on the other hand, are two-dimensional, resulting in a factor of 8 increase in resolution (for half as many particles), which is enough to resolve the smallest pancakes. They confirm that networks of connected pancakes actually form in models such as CDM.

The above argument is useful in explaining the formation of elongated structures even in cosmological models with no cutoff in the power spectrum. However, it does not predict whether these elongated structures will be two-dimensional (pancake-like), with an area of order $\lambda_{NL}^2$, or one-dimensional (filament-like) with a length of order $\lambda_{NL}$. Many authors have addressed the problem of determining if the first structures to collapse are pancakes or filaments, either analytically or numerically, and have reached opposing conclusions. The work by Zel'dovich and his collaborators on the density distribution arising at the nonlinear stage of the gravitational growth of cosmological density fluctuations is reviewed by Shandarin & Zel'dovich (1989). The Zel'dovich approximation (Zel'dovich 1970) describes the linear and early non-linear evolution of the density field by assuming that the total displacement of each mass element is the sum of the individual displacements it would experience if each mode in the power spectrum were acting alone. This approximation predicts that the first structures to form are pancakes. White & Silk (1979) computed analytically the evolution of expanding triaxial ellipsoids. They found that in most cases one axis collapses before the others, forming a flat structure, in agreement with Zel'dovich's predictions. Loeb & Rasio (1994) simulated numerically the gas dynamical evolution of ellipsoids, however, and concluded that such systems tend to collapse into a spindle rather than a flattened structure, in contradiction with the aforementioned results of White & Silk (1979).

Shandarin et al. (1995) performed a series of numerical N-body simulations to investigate the morphology of the *very first* structure that forms, which they define as the locus of points where particle trajectory crossing occurs first. They showed that the first structure that forms is pancake-like, not filament-like. However for their simulations, they used a very restricted range of modes, with no small-scale power. Thus it is not clear whether their results are relevant to models with small-scale power like CDM. The numerical simulations of Little et al. (1991), which used power-law spectra with no small-scale cutoff (a major difference from the simulations of Shandarin et al. [1995]) suggest that filaments dominate over pancakes. However, the presence of small-scale modes breaks up these filaments into clusters even before they have reached nonlinear amplitude. Hence, what they refer to as filaments are actually chains of dense clumps.

Various attempts have been made to characterize the geometry of the nonlinear structure into which matter collapses during the growth of Gaussian-random, linear density fluctuations by studying the statistics of the deformation tensor in the initial conditions (see Doroshkevich 1970; Bardeen et al. 1986). The tendency of nonlinear structures to result from predominantly one-, two-, or three-dimensional flow was shown to be related to the eigenvalues ($\alpha \geq \beta \geq \gamma$) of the deformation tensor associated with the linear density fluctuations. For Gaussian random density fluctuations, the fraction of matter predicted to pass initially through pancake-type caustics (i.e. $\alpha > 0$, $\beta < 0$, $\gamma < 0$) was 42%, while that predicted to collapse along two axes (i.e. $\alpha > 0$, $\beta > 0$, $\gamma < 0$) was 42% and along all three axes (i.e. $\alpha > 0$, $\beta > 0$, $\gamma > 0$) was only 8% (Doroshkevich 1970). Bardeen et al (1986) considered the probability distribution of the ellipticity and prolateness of the density profiles around peaks in the field of Gaussian random linear density fluctuations, filtered on some minimum wavelength scale. They concluded that filamentary structures are more probable than clumps, with pancakes the most probable.

Recently, however, Bond, Kofman, & Pogosyan (1995) applied a new approximate prescription for evolving the linear fluctuations into the nonlinear regime to address the question. They concluded that filamentary structures are present in the initial conditions. Their model predicts that clusters form first, at the location of the highest density peaks, followed by the formation of filaments connecting these clusters, and finally pancakes connecting the filaments.

Bertschinger (1993) and Bertschinger & Jain (1994) developed an original Lagrangian theory based on the evolution of the Weyl tensor, for studying structure formation in pressureless matter, and concluded that gravitational collapse is more likely to form filaments than pancakes. However, they had to assume that the magnetic part of the Weyl tensor is negligible, an assumption that was criticized by Matarrese, Pantano, & Saez (1994). Later, Bertschinger & Hamilton (1994) and Kofman & Pogosyan (1995) showed that the magnetic part of the Weyl tensor is in general non-negligible in the Newtonian limit. Bertschinger & Hamilton (1994) rederived the evolution of the Weyl tensor, and showed that the method does not favor either pancakes or filaments.

Sathyaprakash, Sahni, & Shandarin (1996) performed a percolation analysis of numerical simulation results, and also applied the shape statistics of Babul & Starkman (1992) to their results. They showed that the structures formed in these simulations tend to be more filament-like than pancake-like. They point out, however, that their results do not rule out the possibility that pancakes form first. Finally, various authors (Babul & Starkman 1992; Klypin & Melott 1992; Bertschinger & Hamilton 1994) have given arguments supporting the formation of extended structures in hierarchical models, but do not predict whether filaments or pancakes form first.

In this paper, we will add to this debate by showing that cosmological pancakes are gravitationally unstable with respect to fragmentation within the pancake plane in which matter is concentrated. We shall idealize pancakes as the nonlinear outcome of the growth of a 1D, sinusoidal, plane-wave adiabatic density fluctuation of wavelength $\lambda_p$ in a gas of collisionless matter in an Einstein-de Sitter universe. We will demonstrate that such pancakes become unstable with respect to small amplitude perturbations by modes with $\lambda \lesssim \lambda_p$ and wavevector transverse to that of the pancake mode, shortly after the formation of the nonlinear pancake central layer in which caustics and counterstreaming occur. This result may help to reconcile the apparently contradictory views summarized above in which either pancakes or filaments dominate the process of structure formation in cosmological models. If pancakes are unstable as shown here, that is, then the formation of pancakes can be expected to be accompanied generically by fragmentation and filamentation, even if pancakes form first.

The gravitational instability of thin, 2D concentrations of mass (i.e. surfaces) in both planar and spherical geometry has been considered before, utilizing either linear modal analysis or numerical simulations. Hwang, Vishniac, & Shapiro (1989) studied the gravitational instability of collisionless particles in the cosmological, self-similar, spherical shells which lead the expansion of cosmological underdensities or positive energy perturbations, utilizing a linear modal analysis. They found that such shells are unstable to fragmentation on all scales, and that the growth rate of the perturbation increases logarithmically with the spherical eigennumber $l$ (where $l$ in a spherical shell plays a role similar to that of the wavenumber $\mathbf{k}$, in the sense that the scale length of the perturbation is proportional to $1/l$). White & Ostriker (1990) studied the same problem by numerical N-body simulation. They also found that an isolated cosmologically expanding shell is unstable with respect to nonradial perturbation, and that bound clumps grow within the shell. Anninos, Norman, & Anninos (1995) performed numerical gas dynamical studies of planar pancakes and their fragmentation in 2D, where they imposed initial perturbations with a Poisson noise power spectrum and considered the combined effects of gravity, gas dynamics and radiative cooling on the resulting fragmentation. However, their inclusion of radiative cooling makes it difficult to establish the purely gravitational nature of the instability apart from thermal instability, and makes the results depend upon the pancake wavelength in physical units and the pancake collapse epoch. Likewise, their assumption of Poisson noise for the initial conditions makes it difficult to see the dependence on perturbation wavenumber. Finally, we should note that Alimi et al. (1990) numerically studied nonlinear gravitational collapse of two, orthogonal, plane-wave pancakes of equal amplitude and wavelength, while Moutarde et al. (1995) investigated this problem in 3D for the superposition of three such pancakes, in phase. These simulations show that such conditions lead to filament collapse in 2D and cluster collapse in 3D. These calculations, however, do not address the question of whether cosmological sheets are gravitationally unstable when linearly perturbed, since the superposition of pancake modes of equal amplitude represents a nonlinear mode-mode coupling, instead.

In this paper, we study the gravitational instability of collisionless cosmological pancakes subject to transverse symmetric (density) and antisymmetric (bending, or rippling) perturbation modes by numerical simulations in 2D. We show that nonlinear clumping occurs not only along the line of nodes of two orthogonal pancake modes of equal amplitude and wavelength, but also when a single pancake is perturbed by transverse modes when the perturbation amplitude is much smaller than that of the primary pancake. In particular, we show that when a pancake is perturbed by a transverse mode of small enough amplitude that the transverse mode would, in the absence of the primary pancake mode, remain of linear amplitude until well passed the formation of the primary pancake caustics, the perturbation leads to a *linear* instability of the *nonlinear* pancake, resulting in the fragmentation and, in 2D, the filamentation of the pancake. A similar instability is found if the transverse perturbation amounts to a rippling of the surface of the primary pancake. We shall attempt to answer the following questions by a series of numerical N-body calculations using the

Particle-Mesh method: (1) For what range of wavenumbers are cosmological, 1D plane-wave pancakes gravitationally unstable with respect to transverse perturbations? (2) What is the growth rate of the instability? (3) What is the nonlinear outcome of the instability?

In §2, we describe our calculations. This includes our basic equations, the background flow of the unperturbed pancake, the perturbation modes, our numerical method, and the range of cases and parameters considered. In §3, we discuss the gravitational instability of cosmological sheets utilizing a thin-sheet energy argument. In §4, we present our simulation results, demonstrating and quantifying the instability, its growth rate, and the nonlinear outcome, including a brief description of the filamentary structures formed as the result of instability. In §5, we give our summary and conclusions.

## 2. THE CALCULATION

### 2.1. Basic Equations

We assume that the matter density in the universe is dominated by a single component of collisionless matter and focus our attention on the evolution of that component in a matter-dominated Friedmann universe. We have followed Shandarin (1980) in defining a set of dimensionless comoving variables as follows:

$$\tilde{r}_i = a^{-1}(r_i/r_0), \tag{2}$$
$$d\tilde{t} = a^{-2}(dt/t_0), \tag{3}$$
$$\tilde{\rho} = a^3(\rho/\rho_0), \tag{4}$$
$$\tilde{v}_i = a(v_i/v_0), \tag{5}$$

and

$$\tilde{\phi} = a^2(\phi/\phi_0), \tag{6}$$

where the cosmic scale factor $a = (1+z)^{-1} = \tilde{t}^{-2}$ in the Einstein-de Sitter case ($\Omega_0 = 1$) considered here, $z$ is the redshift, $t$ is the proper time, $r_i$ are the proper spatial coordinates, $\rho$ is mass density, $\phi$ and $v_i$ are the "peculiar" gravitational potential and velocity, respectively, and all zero-subscripted variables are fiducial values of corresponding physical quantities. In particular, these fiducial values are fixed by the value of the Hubble constant $H_0$ [the present value of $H(t)$] and by our choice of a value of the parameter $r_0$ with dimensions of length, as follows:

$$t_0 = \frac{2}{H_0}, \tag{7}$$

$$v_0 = \frac{r_0}{t_0}, \tag{8}$$

$$\rho_0 = \frac{3H_0^2}{8\pi G}, \tag{9}$$

and

$$\phi_0 = v_0^2. \tag{10}$$

The collisionless matter component is described in the Newtonian approximation by the Vlasov equation for the evolution of the single-particle phase space distribution function $f(t, r, v)$ in "tilde" variables, given by

$$\frac{\partial \tilde{f}}{\partial \tilde{t}} + (\mathbf{v} \cdot \nabla)\tilde{f} - \nabla \tilde{\phi} \cdot \nabla_{\tilde{\mathbf{v}}} \tilde{f} = 0. \tag{11}$$

The peculiar gravitational potential of these collisionless particles is then given by the Poisson equation in tilde coordinates as

$$\nabla^2 \tilde{\phi} = 6a(\tilde{t})(\tilde{\rho} - 1), \tag{12}$$

where

$$\tilde{\rho} = \int \tilde{f} d^3\tilde{\mathbf{v}}. \tag{13}$$

### 2.2. 1D Plane Wave Pancake: The Unperturbed Background Flow

Our unperturbed background flow is the growing mode of a sinusoidal, adiabatic, plane wave, cosmological density fluctuation, in an expanding Einstein-de Sitter universe. The equations for initial position, velocity, density and peculiar gravitational potential are given by:

$$\tilde{x} = \tilde{q}_x + \frac{\delta_i}{2\pi \tilde{k}_p} \sin 2\pi \tilde{k}_p \tilde{q}_x, \tag{14}$$

$$\tilde{y} = \tilde{q}_y, \tag{15}$$

$$\tilde{v}_x = \frac{\dot{\delta}_i}{2\pi \tilde{k}_p} \sin 2\pi \tilde{k}_p \tilde{q}_x, \tag{16}$$

$$\tilde{v}_y = 0, \tag{17}$$

$$\tilde{\rho} = \frac{\bar{\tilde{\rho}}}{1 + \delta_i \cos 2\pi \tilde{k}_p \tilde{q}_x}, \tag{18}$$

and

$$\tilde{\bar{\phi}} = \bar{\tilde{\phi}}_i \cos 2\pi \tilde{k}_p \tilde{x}, \qquad (19)$$

where $\tilde{k}_p = 1/\tilde{\lambda}_p$, $\tilde{\lambda}_p = \lambda_p/r_0$, $\delta(\tilde{t}) = a(\tilde{t})/a_c$, $\delta_i = \delta(\tilde{t}_i)$, $\dot{\delta}(\tilde{t}) = 2a^{3/2}/a_c$, $\dot{\delta}_i = \dot{\delta}(\tilde{t}_i)$, $\tilde{\bar{\rho}} = 1$, $\bar{\tilde{\phi}} = 6\delta_i a_i/(2\pi \tilde{k}_p)^2$, and $\lambda_p$ is the present comoving value of the pancake wavelength in physical units (see, for example, Shapiro, Struck-Marcell, & Melott [1983]; Shapiro & Struck-Marcell [1985]). Notice that $a_c$ is the scale factor at which a density caustic first forms along the central plane of the initial density maximum (in this case at $\tilde{x} = \tilde{q}_x = 0.5$). This solution is independent of the wavelength $\lambda_p$ of the plane wave and of the initial amplitude, $\delta_i$, when lengths are expressed in units of $\tilde{\lambda}_p$ and the time is expressed in terms of $a(\tilde{t})/a_c$. From this point on, for simplicity of notation, we drop the tilde from our variables, but unless otherwise noted, all our quantities are in tilde variables as defined in §2.1.

### 2.3. Perturbations

We perturb the background solution described above with one of two distinct modes described as follows.

#### 2.3.1. Symmetric (Density) Mode

In this mode, a transverse plane wave density fluctuation with wavenumber $k_s = \lambda_p/\lambda_s$ (where $\lambda_s$ is the symmetric perturbation wavelength), and amplitude $\epsilon_s \delta_i$ is added to the unperturbed pancake background flow. The initial position, velocity, density and peculiar gravitational potential are then given by

$$x = q_x + \frac{\delta_i}{2\pi k_p} \sin 2\pi k_p q_x, \qquad (20)$$

$$y = q_y + \epsilon_s \frac{\delta_i}{2\pi k_s} \sin 2\pi k_s q_y, \qquad (21)$$

$$v_x = \frac{\dot{\delta}_i}{2\pi k_p} \sin 2\pi k_p q_x, \qquad (22)$$

$$v_y = \epsilon_s \frac{\dot{\delta}_i}{2\pi k_s} \sin 2\pi k_s q_y, \qquad (23)$$

$$\rho = \frac{\bar{\rho}}{1 + \delta_i (\cos 2\pi k_p q_x + \epsilon_s \cos 2\pi k_s q_y)}, \qquad (24)$$

and

$$\phi = \bar{\phi}(\cos 2\pi k_p x)\left(1 + \epsilon_s \frac{k_p^2}{k_s^2} \frac{\cos 2\pi k_s y}{\cos 2\pi k_p x}\right) \qquad (25)$$

where both $\delta_i \ll 1$ and $\epsilon_s \ll 1$, $\bar{\phi}$ is defined as in the previous section, and $\bar{\rho} = 1$ again. The maximum of this spatially varying density perturbation is given by

$$\left|\frac{\delta\rho}{\bar{\rho}}\right|_{\max} = \delta_i(1 + \epsilon_s), \tag{26}$$

which occurs at $x = n/2k_p$ and $y = n/2k_s$ where $n = 1, 3, ...$ for $0 \leq y \leq 1$ and $0 \leq x \leq 1$.

### 2.3.2. Antisymmetric (Bending or Rippling) Mode

In this mode, the unperturbed plane of symmetry of the pancake is rippled by imposing a small, antisymmetric, sinusoidal, transverse perturbation of wavenumber $\mathbf{k}_a = k_a \hat{y}$, where $k_a = \lambda_p/\lambda_a$ (where $\lambda_a$ is the antisymmetric perturbation wavelength) and amplitude $\epsilon_a$ on the peculiar gravitational potential fluctuation of the initial, unperturbed pancake, given by

$$\phi = \bar{\phi}\cos\left[(2\pi k_p)\left(x + \frac{\epsilon_a}{2\pi k_a}\frac{k_p}{k_a}\cos 2\pi k_a y\right)\right], \tag{27}$$

where $\epsilon_a \ll 1$. (Our choice of normalization to render $\epsilon_a$ dimensionless derives from the desire to obtain a similar expression for the gravitational potential as in the symmetric case to simplify the comparisons of amplitudes). Alternatively, we can expand the cosine term in the above equation and find to linear order in $\epsilon_a$,

$$\phi = \bar{\phi}(\cos 2\pi k_p x)\left(1 - \epsilon_a \frac{k_p^2}{k_a^2}\frac{\cos 2\pi k_a y \sin 2\pi k_p x}{\cos 2\pi k_p x}\right). \tag{28}$$

In this form it is easily seen that the perturbation amplitude $\epsilon_a$ plays a role analogous to that of $\epsilon_s$ for the symmetric mode (see eq. [25]).

The initial position and velocity are given for this case by solving the momentum equation, yielding (to linear order):

$$x = q_x + \frac{\delta_i}{2\pi k_p}S_x, \tag{29}$$

$$y = q_y - \frac{\delta_i}{2\pi k_a}\epsilon_a S_x S_y, \tag{30}$$

$$v_x = \frac{\dot{\delta}_i}{2\pi k_p}S_x, \tag{31}$$

and

$$v_y = -\frac{\dot{\delta}_i}{2\pi k_a}\epsilon_a S_x S_y \,. \tag{32}$$

The initial density to second order in $\epsilon_a$ is given from the continuity equation by

$$\rho = \bar{\rho}\Big[1 + \delta_i C_x - \delta_i \epsilon(S_x C_y - 2\pi k_p \epsilon C_x S_y^2)\Big]^{-1} \,, \tag{33}$$

where

$$\chi \equiv x + \frac{\epsilon_a}{2\pi k_a}\frac{k_p}{k_a}\cos 2\pi k_a y \,, \tag{34}$$

$$C_y \equiv \cos 2\pi k_a y \,, \tag{35}$$

$$C_x \equiv \cos 2\pi k_p \chi \,, \tag{36}$$

$$S_y \equiv \sin 2\pi k_a y \,, \tag{37}$$

and

$$S_x \equiv \sin 2\pi k_p \chi \,. \tag{38}$$

These initial conditions are vorticity-free, consistent with the fact that the gravitational growth of initially vorticity-free, primordial density fluctuations keeps them vorticity-free prior to the occurrence of caustics, or shell-crossing. The maximum value of this spatially varying density perturbation is given by

$$\left|\frac{\delta\rho}{\bar{\rho}}\right|_{\max} = \delta_i\left(1 + \epsilon_a^2\right)^{1/2} \,, \tag{39}$$

which is of second order in $\epsilon_a$. Comparison of equations (26) and (39) shows that for the same $\epsilon$ and initial amplitude $\delta_i$, the symmetric mode has a larger maximum spatially varying density perturbation, $|\delta\rho/\bar{\rho}|_{\max}$, than the antisymmetric mode.

### 2.4. Numerical Method

The gravitational dynamics of this collisionless matter in an Einstein-de Sitter universe is modeled by a numerical N-body code based upon the Particle-Mesh (PM) method in 2D, Cartesian $(x, y)$ coordinates (see Villumsen 1989; Shapiro et al. 1996). The Poisson equation is solved on a uniform, square grid with Fast Fourier Transform techniques and periodic boundary conditions. The mass is assigned to the grid using the cloud-in-cell scheme (Hockney & Eastwood 1981). The equations of motion are integrated with a second order Runga-Kutta scheme (Shapiro & Martel 1996).

## 2.5. Growth Rate Calculations

To quantify the growth of instability and its dependence on wavenumber, we shall perform a Fourier analysis of the density field. We calculate the surface density $\Sigma$ of all the mass within a distance $\lambda_p/2$ on both sides of the unperturbed pancake central plane (defined above) as a function of the transverse coordinate $y$, according to

$$\Sigma(y) = \int_0^1 \rho(x,y)dx \,. \tag{40}$$

The surface density fluctuation $\delta_\Sigma$ is then just

$$\delta_\Sigma(y) \equiv \frac{\Sigma(y) - \bar{\Sigma}}{\bar{\Sigma}} \,. \tag{41}$$

It can also be written in terms of its inverse discrete Fourier transform $\hat{\delta}_\Sigma(k)$ as

$$\delta_\Sigma(y) = \sum_k \hat{\delta}_\Sigma(k) e^{-2\pi i k y} \,. \tag{42}$$

The power spectrum $P_\Sigma(k)$ of the surface density fluctuation can then be found from

$$P_\Sigma(k) = \langle |\hat{\delta}_\Sigma(k)|^2 \rangle \,, \tag{43}$$

once we solve for $\hat{\delta}_\Sigma(k)$ from equation (42) above. We define a dimensionless growth rate $\tilde{\Gamma}$, calculated from the power spectrum, which depends on time only as a function of $a/a_c$, according to

$$\tilde{\Gamma} = \Gamma t = \left[\frac{\dot{\delta}_\Sigma(k)}{\delta_\Sigma(k)}\right] t = \frac{1}{3}\left(\frac{a}{a_c}\right)\frac{d \ln P_\Sigma(k)}{d(a/a_c)} \,, \tag{44}$$

where $t$ is the proper Hubble time at epoch $a/a_c$, not the $\tilde{t}$ time, and we have used the relation $t/t_c = (a/a_c)^{3/2}$; $\dot{\delta}_\Sigma$ denotes the derivative of $\delta_\Sigma$ with respect to $t$. (Note that the normalization of the power spectrum is not important for growth rate calculations, since only the derivative of its logarithm appears in the growth rate equation.) For example, if pancakes prove to be gravitationally unstable and the instability grows as a power law in time, according to $\delta_\Sigma \propto t^\alpha$, then $\tilde{\Gamma} = \Gamma t = \alpha = $ constant will be found.

## 2.6. Epoch of Nonlinearity

We wish to estimate the epoch at which the transverse symmetric or antisymmetric perturbations become nonlinear and measure the level of nonlinearity generated in the clusters that form as a result of the instability. We calculate the density using the well-known cloud in cell (CIC) scheme used in our PM method gravity solver, on a grid with a mesh size equal to the mean particle spacing $\Delta = L/N_{p,1}$, where $L$ is the box size, equal to unity in computational units, and $N_{p,1}$ is the mean number of particles per dimension, where the total number of particles in the box is $N_{p,1}^2$. We then calculate $\delta\rho/\bar{\rho} = (\rho - \bar{\rho})/\bar{\rho}$ on the grid (where $\bar{\rho}$ is the mean density in the computational box; in tilde units, $\bar{\rho} = 1$). We estimate a measure of nonlinearity for the entire system by calculating the mass-weighted, rms density fluctuation $\sigma_\rho$, given by

$$\sigma_\rho^2 = \frac{\int \rho(\delta\rho/\bar{\rho})^2 dxdy}{\int \rho \, dxdy}, \qquad (45)$$

and normalize it by dividing it by $\sigma_{\rho,0}^2$ where $\sigma_{\rho,0}^2 = \sigma_\rho^2$ (unperturbed pancake). Perturbations of the primary pancake become nonlinear when $\sigma_n^2 = \sigma_\rho^2/\sigma_{\rho,0}^2 \geq 2$.

## 2.7. Cases and Initial Parameters

We have performed 2D, PM simulations of purely collisionless pancakes, perturbed by either symmetric or antisymmetric modes with wavevectors transverse to the direction of pancake collapse, for a range of initial wavenumbers $k$, and amplitudes $\epsilon$. The perturbation wavenumbers that we simulated were 1/4, 1/2, 1, 2, 4, 8, 16, 32, and 64. For the symmetric mode, we simulated cases with perturbation amplitudes 0.001, 0.01, 0.05, 0.1, and 0.2. The perturbation amplitudes for cases of antisymmetric mode were 0.05, 0.1 and 0.2 (and, for $k_a = 64$, $\epsilon_a = 0.01$, as well). We have labeled each case according to the perturbation mode (S for Symmetric and A for Antisymmetric). The first subscript of the letter S or A denotes the value of the perturbation wavenumber $k$ and the second denotes the value of the initial perturbation amplitude, $\epsilon$. For example, case $A_{2,0.1}$ corresponds to an antisymmetric mode with wavenumber $k_a = 2$ and amplitude $\epsilon_a = 0.1$, where, in our dimensionless units, the wavelength of the primary pancake is equal to unity (i.e. $r_0 = \lambda_p$ and, hence $\tilde{\lambda}_p = 1$). Notice that for the cases with $k < 1$, our assumption of periodic boundary conditions requires us to take $r_o > \lambda_p$. For example, for $k = 1/2$, $r_0 = 2\lambda_p$ and $\tilde{\lambda}_p = 1/2$, while for $k = 1/4$, $r_0 = 4\lambda_p$ and $\tilde{\lambda}_p = 1/4$. The initial amplitude, $\delta_i = a_i/a_c = 0.15$, was chosen such that the primary pancake mode, in the absence of transverse perturbations, would collapse at an epoch $a/a_c = 1$. Simulations were terminated at $a/a_c = 7$. We used $256^2$ particles on $512^2$ gravity cells for cases with perturbation wavenumber less than or equal to 8. We used $512^2$ particles on $1024^2$ gravity cells for cases with perturbation wavenumber greater than 8.

## 3. GRAVITATIONAL INSTABILITY: THIN SHEET APPROXIMATION AND THE ENERGY ARGUMENT

### 3.1. Thin Sheet Stability Analysis

The gravitational instability of the central layer of the pancake – the region in which density caustics and counterstreams occur in the collisionless particles, bounded by two, symmetrically placed accretion shocks when gas is present – can be estimated by a thin sheet, energy argument similar to the approximation made by Ostriker & Cowie (1981). We approximate the collisionless pancake central layer as an infinite, plane sheet of mass of zero thickness. The total energy per unit mass, $E(t)$, of a circular disk of diameter $\lambda$ cut out of this infinitesimally thin, collisionless pancake layer of total surface density $\Sigma(t)$ at time $t$, expanding uniformly in its plane with a Hubble law characterized by a Hubble constant $H(t)$, is given by

$$E = \frac{1}{16}H^2\lambda^2 - \pi G K \Sigma \frac{\lambda}{2} + \frac{1}{2}v_{\rm rms}^2, \qquad (46)$$

where $K = 0.849$ for a flat disk. Here and throughout §3 we express all quantities in physical units, rather than the tilde variables defined in §2 unless otherwise noted. The terms on the right hand side represent cosmological expansion energy, gravitational potential energy, and internal energy, respectively, for the collisionless particles inside the pancake. In principle, there are two sources of velocity dispersion inside the pancake central layer, that due to the random thermal motions present initially in the unperturbed gas and that due to the streaming motion induced by the plane-wave perturbation which formed the pancake. We let $\langle v_{\rm th}^2 \rangle^{1/2}$ refer to rms velocity dispersion associated with the finite initial "temperature," while $\langle v_{\rm s}^2 \rangle^{1/2}$ refers to the rms "vertical velocity dispersion" (i.e. that due to the infall and outflow of particles along the $x$-direction, perpendicular to the pancake plane). In general, $\langle v_{\rm rms}^2 \rangle^{1/2} = \langle v_{\rm th}^2 \rangle^{1/2} + \langle v_{\rm s}^2 \rangle^{1/2}$ in equation (46) above. For pancakes in a collisionless gas which is initially "cold," $\langle v_{\rm th}^2 \rangle^{1/2} = 0$, and $v_{\rm rms}$ is due entirely to "vertical velocity dispersion" $\langle v_{\rm s}^2 \rangle^{1/2}$, associated with pancake collapse.

According to the thin sheet approximation, bound collisionless fragments can form only if $E$ is negative. According to equation (46), this occurs for a finite range of values of $\lambda$. Expressed in terms of the dimensionless wavenumber, $k = \lambda_p/\lambda$ (where $\lambda_p(t) = \lambda_{p,0} a$ and $\lambda_{p,0}$ is the present comoving value), equation (46) implies that a transverse perturbation is gravitationally unstable if $k_{\rm min} < k < k_{\rm max}$, where $k_{\rm min}$ is the critical wavenumber below which Hubble expansion stabilizes the sheet against gravitational fragmentation, and $k_{\rm max}$ is that above which the internal kinetic energy associated with vertical motions and finite temperature is the stabilizing agent. The maximum growth rate corresponding to $k = k_{\Gamma_{\rm max}}$, is found by defining a growth rate $\Gamma$ such that

$$\Gamma^2 \equiv -E k^2 \lambda_p^{-2} \,, \tag{47}$$

and requiring that $\partial \Gamma / \partial k |_{k_{\Gamma_{\max}}} = 0$. This yields

$$k_{\Gamma_{\max}} = \frac{\pi G K \Sigma \lambda_p}{2 v_{\rm rms}^2} \,. \tag{48}$$

We note that this result is analogous to those derived from linear analysis in the case of a cosmological expanding shell comprised of gaseous baryonic matter (Vishniac 1983b) for which $\lambda_{\Gamma_{\max}} = (\pi G \Sigma / c_s^2)^{-1}$. In terms of the wavevector of maximum growth rate, the critical wavenumbers $k_{\min}$ and $k_{\max}$ can be expressed as follows:

$$k_{\min} \ (\text{or} \ k_{\max}) = k_1 \left\{ 1 \pm \left[ 1 - \frac{1}{8} \left( \frac{v_{H,\max}}{v_{\rm rms}} \right)^2 \right]^{1/2} \right\}^{-1} \,, \tag{49}$$

where

$$k_1 \equiv \frac{\lambda_p H^2}{4 \pi G K \Sigma} \,, \tag{50}$$

and

$$v_{H,\max} \equiv \frac{\lambda_p H}{k_{\Gamma_{\max}}} \,, \tag{51}$$

where $k_{\min}$ ($k_{\max}$) corresponds to the plus (minus) sign in equation (49) above.

## 3.2. Results of Thin Sheet Stability Analysis Applied to 1D, Collisionless, Zel'dovich Pancake

To estimate $k_{\min}$, $k_{\max}$, and $k_{\Gamma_{\max}}$ numerically as a function of $a/a_c$, we have used a one-dimensional Lagrangian hydrodynamics-plus-N-body code (Shapiro & Struck-Marcell 1985) which includes a collisionless particle component to simulate a 1D pancake in the limit where $\Omega_{\rm gas}$ goes to zero (i.e. $\Omega_{\rm dm} = 1$). Our simulation took as its initial conditions those of the unperturbed background flow described in §2.1. Results for one time-slice, $a/a_c = 3.5$, are shown plotted in Figure 1. The "thin sheet" in this case is defined as the matter concentrated along the central plane of the pancake, bounded by the outermost density caustics, one on either side of the central plane. If a collisional gas component is also present, these outermost caustics in the collisionless component are roughly spatially coincident with the strong accretion shocks of the collisional gas component, if that component has a ratio of specific heats $\gamma = 5/3$ and the postshock gas is adiabatic. We note that, in this calculation, $v_{rms}$ is calculated from the speed of the collisionless particles moving inside the pancake layer with a vertical streaming velocity $v_s$ in one dimension.

The minimum and maximum wavenumbers for which the pancake is unstable to perturbation and the wavenumber which has the fastest growth rate are shown as a function of $a/a_c$ in Figure 2. The region between $k_{min}$ and $k_{max}$ signifies the unstable region. The pancake first becomes unstable shortly before $a/a_c \sim 1.5$, and, thereafter, the unstable region grows. We have calculated the dimensionless growth rate, $\tilde{\Gamma} = \Gamma t$, as a function of $a/a_c$ for several perturbation modes according to equation (47). The results are plotted in Figure 3 for unstable wavenumbers $k = 1, 2, 4, 8$ and $k_{\Gamma_{max}}$. For $k \lesssim k_{\Gamma_{max}}$, the growth rate varies with $k$ roughly as $k^{1/2}$, as expected from equation (47). However this relation breaks down for $k \sim k_{min}$ or $k \gtrsim k_{\Gamma_{max}}$.

## 4. SIMULATION RESULTS AND ANALYSIS

### 4.1. Qualitative Overview

Our 2D simulations indicate that pancakes are gravitationally unstable with respect to transverse perturbation modes of either the symmetric or antisymmetric kind for all perturbation wavenumbers $k \gtrsim 1$. In the discussion below, we first describe the qualitative effects of perturbations by these unstable modes. We then quantify these effects by calculating the power spectrum of the fluctuations of pancake surface density, presenting growth rates of the instability, and identifying the onset of nonlinearity. We compare and contrast these results with the predictions of the thin-sheet energy argument. Finally, the density profiles of nonlinear clumps (i.e. filaments in 2D) are described.

#### 4.1.1. Symmetric Mode

As an illustration of the qualitative outcome of symmetric mode perturbations, we focus on the case $k_s = 2$. Figures 4a and 4b show the particle positions at $a/a_c = 1$ and $a/a_c = 7$, respectively, for our 2D, PM simulations of a pancake perturbed by a mode $S_{2,0,2}$. These figures indicate that, for unstable $k_s$, the symmetric mode generally leads to dense clumps in the central plane, at the intersection of the nodes of transverse density perturbation with the central plane (one per $\lambda_s = 1/k_s$). In our 2D simulations, these clumps are actually filaments viewed edge-on in cross-section.

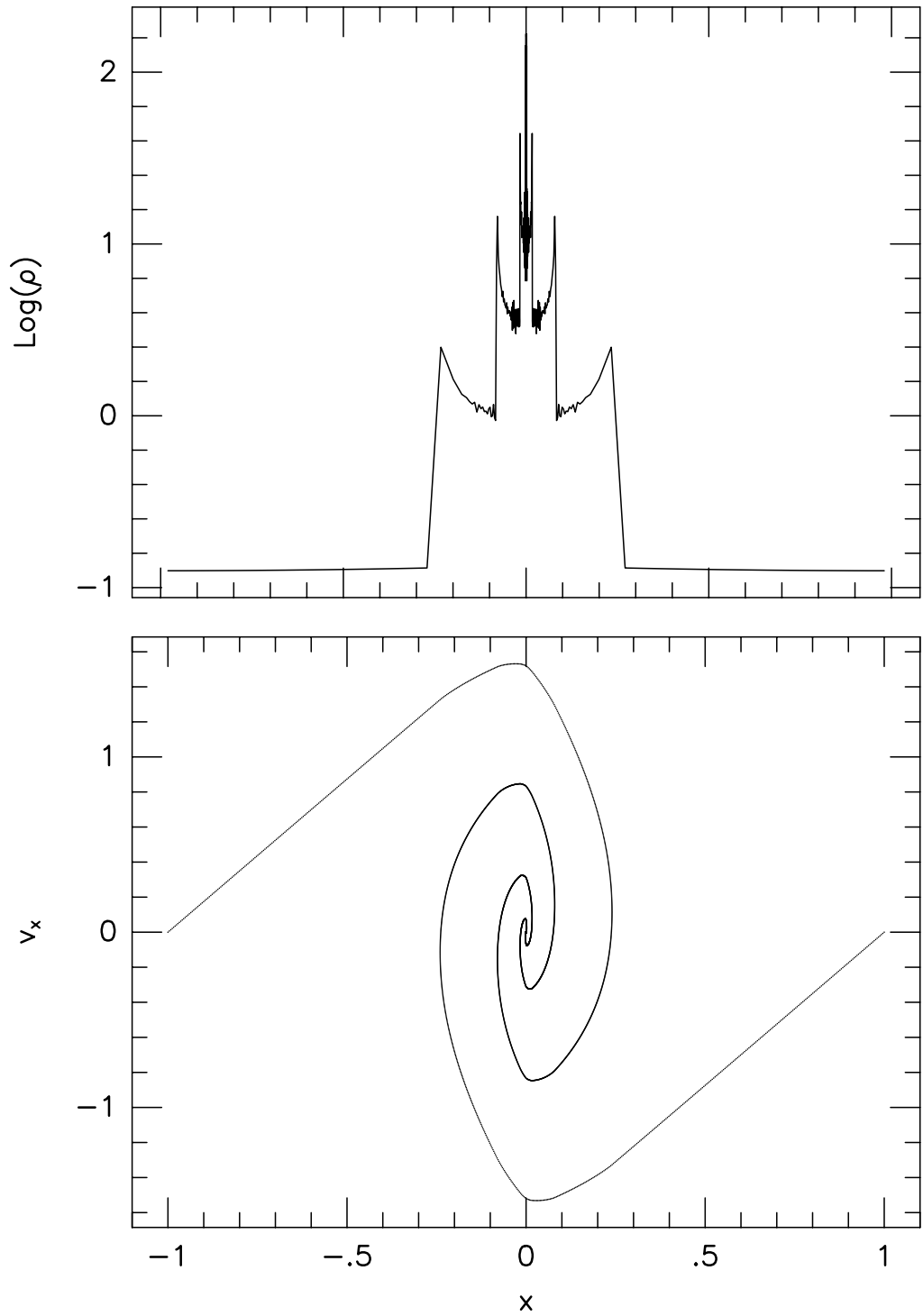

Fig. 1.— Unperturbed, 1D Pancake Simulation Results at $a/a_c = 7.0$. Density $\tilde{\rho}$ (top panel) and velocity $\tilde{v}_x$ (bottom panel) are plotted versus $\tilde{x}$, the distance from the pancake midplane. All quantities are in tilde variables with $r_0 = \lambda_p$.

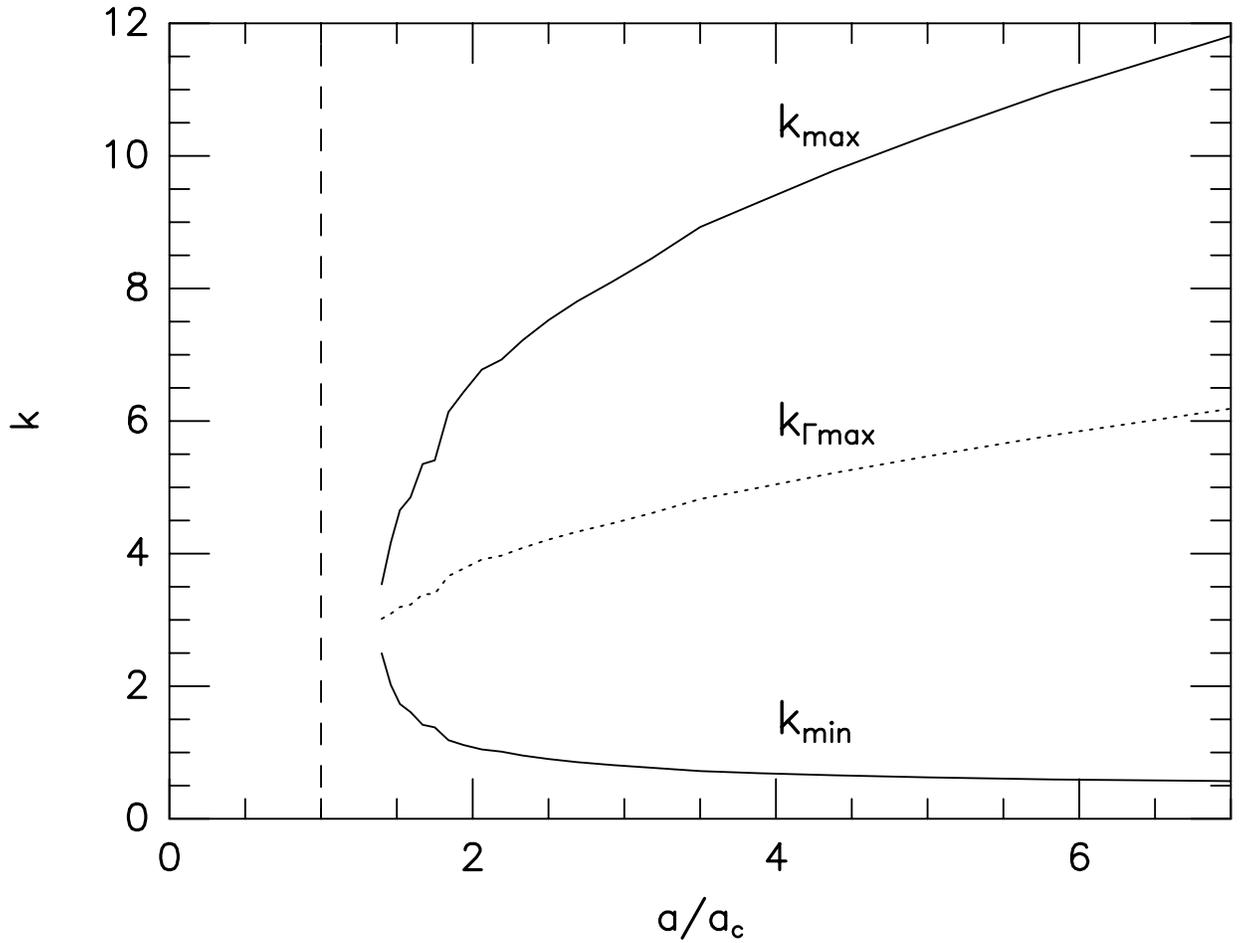

Fig. 2.— Unstable wavenumbers $k$ vs. scale factor $a/a_c$ according to the thin-sheet energy argument applied to 1D pancake simulation results. The solid lines trace the boundary of the unstable region. The dotted line signifies the wavenumber $k_{\Gamma_{max}}$ which has the fastest growth rate. The vertical dashed line indicates the epoch of caustic formation for the pancake.

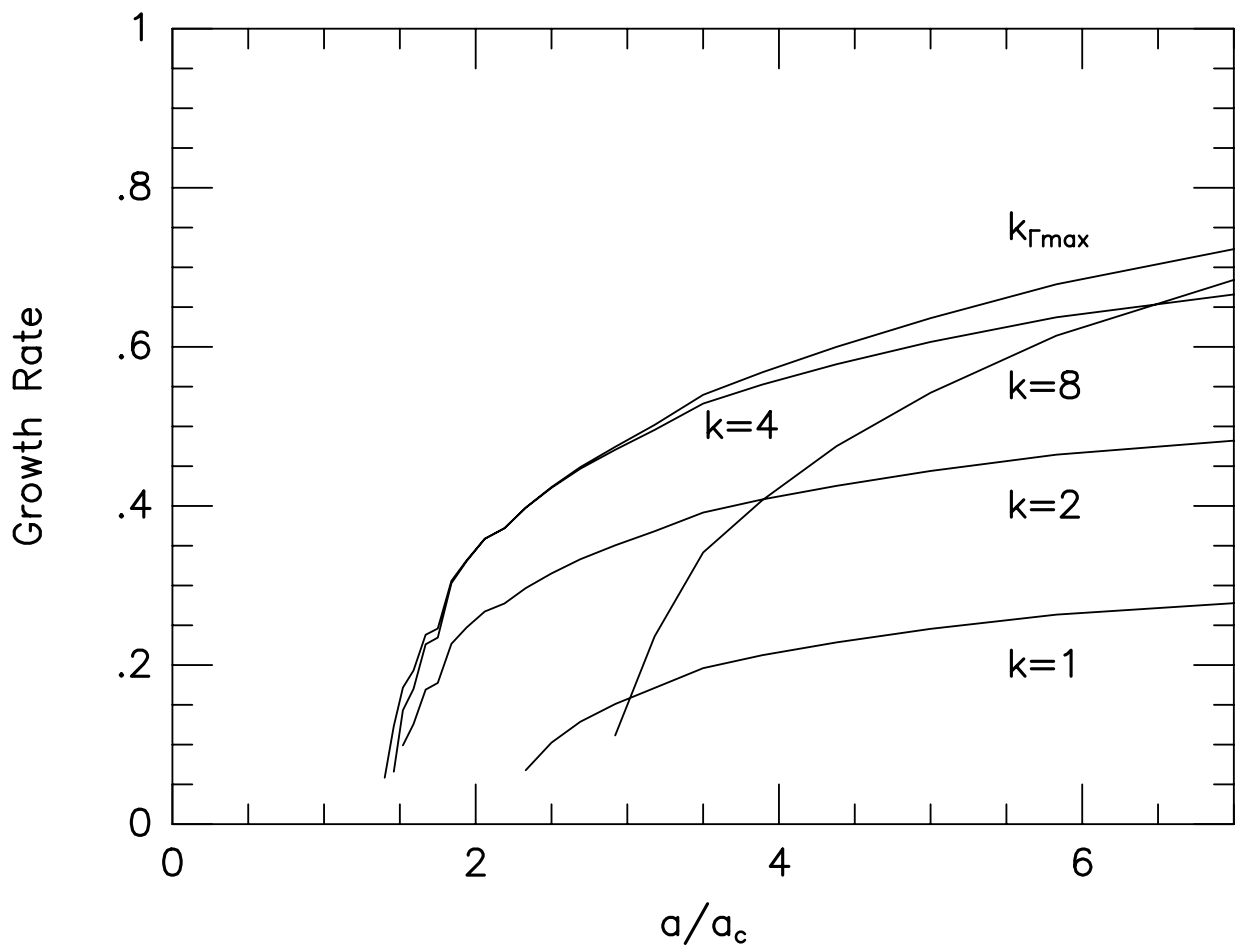

Fig. 3.— (a) (this page) Dimensionless growth rate $\tilde{\Gamma}$ of gravitational instability in pancake vs. $a/a_c$, calculated from the thin-sheet energy argument applied to 1D pancake simulation results. All curves are labelled with the value of the perturbation wavenumber, except the uppermost curve, which corresponds to the maximum growth rate, that for mode $k_{\Gamma_{max}}$ (as illustrated in Fig. 2). (b) (next page) Dimensionless growth rate $\tilde{\Gamma}$ vs. perturbation wavenumber $k$ at fixed time slices. Each curve is labelled with the value of $a/a_c$.

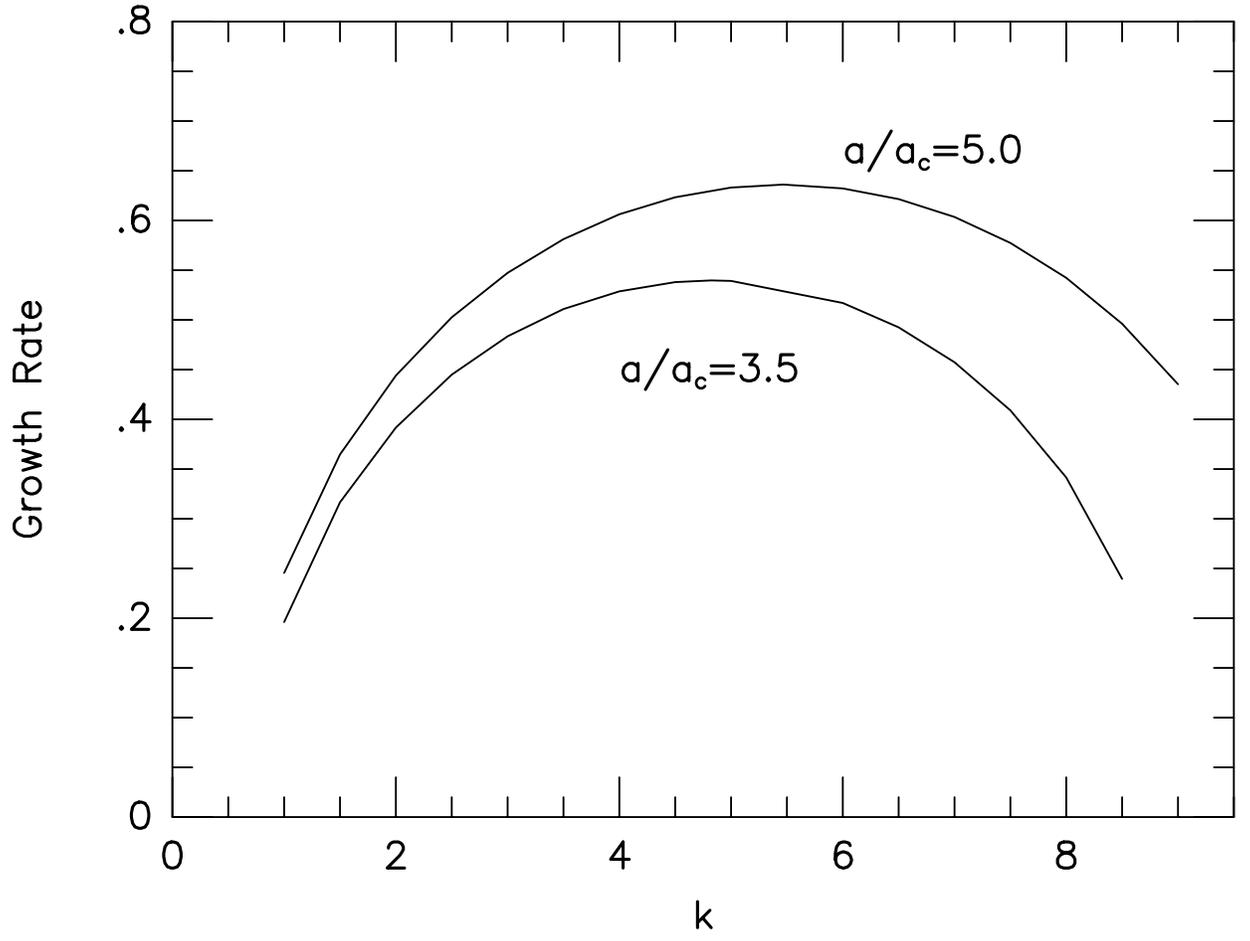

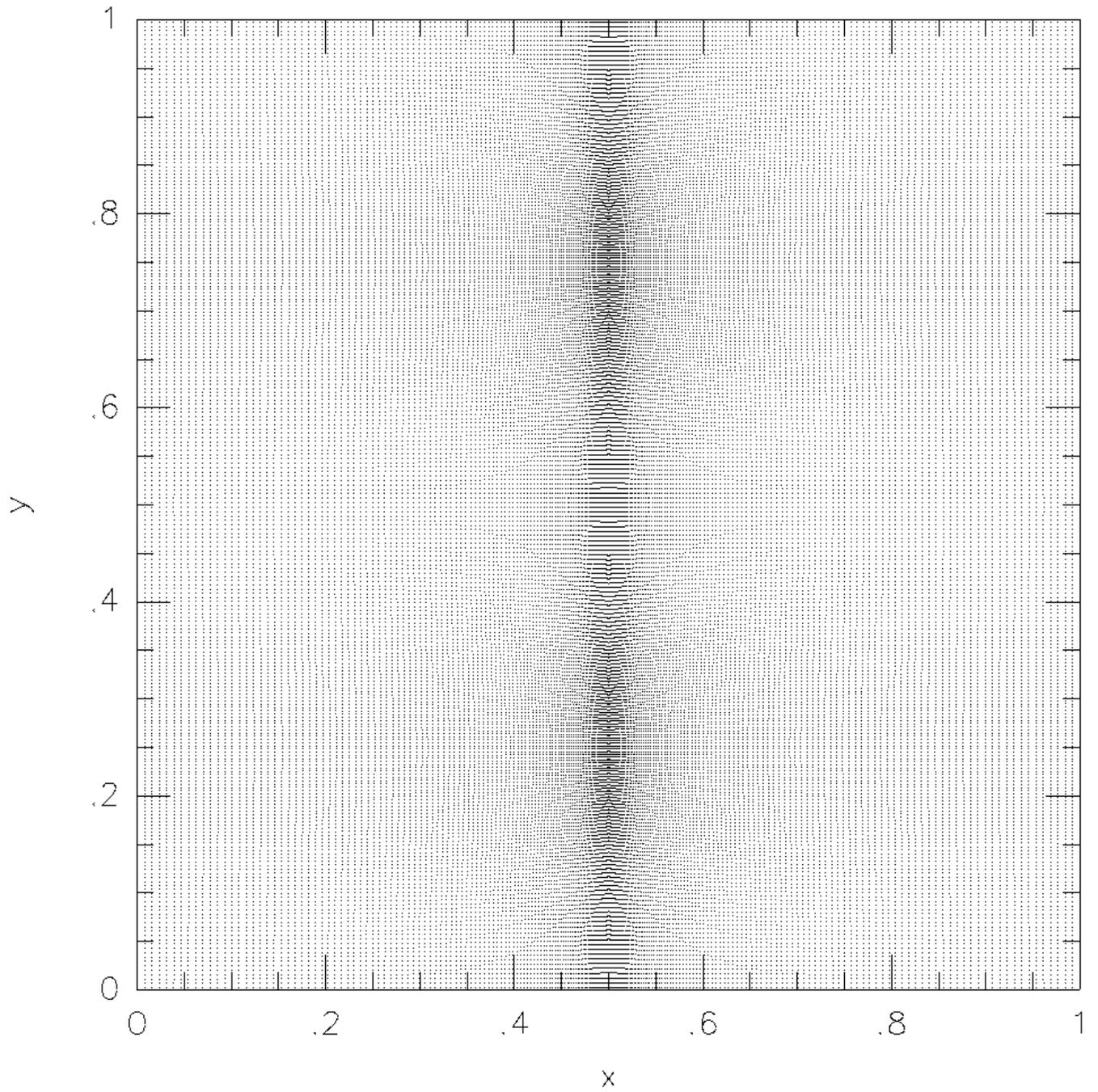

Fig. 4.— (a) (this page) Symmetric Mode. Particle positions at $a/a_c = 1$ for a PM simulation of a pancake perturbed by mode $S_{2,0,2}$. (b) (next page) Same as (a), but for $a/a_c = 7$.

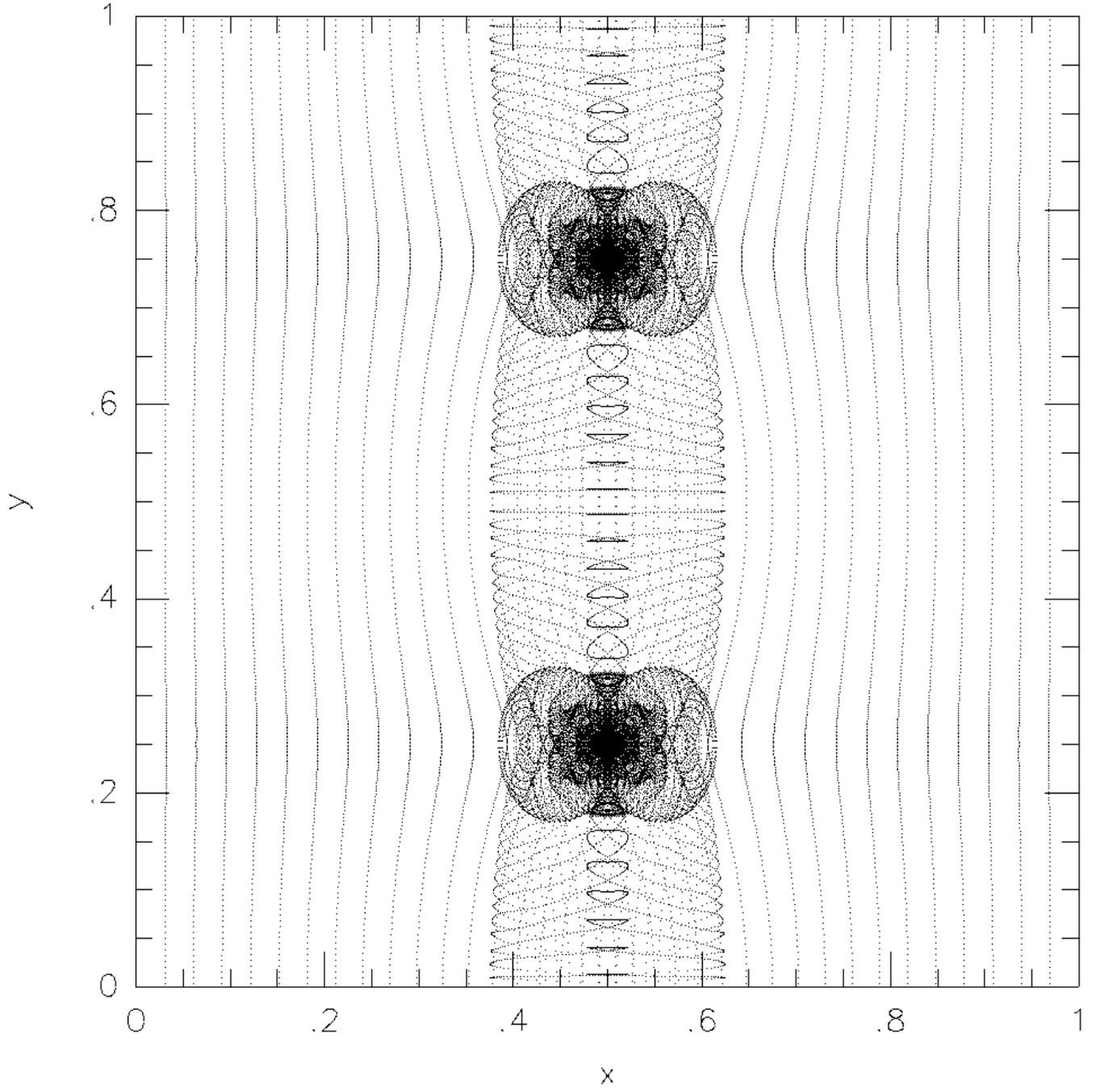

As we shall see, these results indicate that the nonlinear pancake central layer is *linearly* unstable with respect to transverse perturbation. However, before we convince the reader of this in the following sections, we note that some of the behavior of the perturbed pancake can be understood as a simple consequence of the *linear* superposition of two transverse pancake modes of unequal amplitudes and wavelengths. From linear analysis we expect that, in the absence of the transverse perturbation mode, the primary pancake will form density caustics for the first time at epoch $a/a_c = 1$, while, in the absence of the primary pancake, the transverse perturbation will caustic at $a/a_c = 1/\epsilon_s = 5$. However, in the presence of the transverse mode, we expect the primary mode to become nonlinear sooner than the predicted caustic time for the unperturbed pancake. Due to the superposition of the two orthogonal modes, the effective linear amplitude of the primary pancake at the points where the central plane intersects the nodes of the transverse perturbation is larger before the formation of nonlinearity than without the transverse mode. This effective linear amplitude is given by

$$\delta_{\text{eff}} = (\delta_{i,1} + \delta_{i,2}) \frac{a}{a_i}, \qquad (52)$$

where $\delta_{i,2} = \epsilon_s \delta_{i,1}$. According to equation (52), the primary pancake is expected to reach nonlinear amplitude at the locations of the nodes in the transverse perturbation (i.e. $\delta_{\text{eff}} \cong 1$) at $a/a_i = (a_{c,1+2}/a_i)$ given by

$$\frac{a_{c,1+2}}{a_i} \cong \left(\frac{\delta_{i,1}}{\delta_{i,1} + \delta_{i,2}}\right)\left(\frac{a_{c,1}}{a_i}\right), \qquad (53)$$

where $a_{c,1}$ is the scale factor at which the primary pancake density caustics first form if the transverse perturbation is absent. For $\delta_{i,2} = \epsilon_s \delta_{i,1}$, this yields $a_{c,1+2}/a_i \cong (1 + \epsilon_s)^{-1}(a_{c,1}/a_i)$. In the example shown in Figures 4a and 4b, therefore, $a_{c,1+2}/a_i \cong 0.83$, while $a_{c,1}/a_i = 1$.

We also expect an enhancement of the linear growth rate of the transverse pancake (i.e. the transverse symmetric mode) in the presence of the emergence of nonlinearity due to the primary pancake. One way to see this is in terms of the enhancement of the linear growth rate of the transverse pancake which would occur if the transverse pancake occurred in a closed universe (i.e. $\Omega_{\text{eff}} = \rho_{\text{pancake}} > 1$). Inside the overdense primary pancake central layer, the transverse mode grows somewhat faster than it does outside the central layer, as if it is in a denser universe there. Due to the nonlinear nature of the problem, however, an exact estimate of the epoch at which the transverse pancake forms a density caustic in the central plane of the primary pancake requires a numerical study and is outside the scope of our discussion.

In this paper, however, we would like to distinguish between the nonlinearity that forms at the intersection of two pancakes with equal perturbation amplitudes and orthogonal wavevectors and the nonlinearity that will form as a result of the superposition of two orthogonal density fluctuations when the amplitude of one is much smaller than the amplitude of the other. In the former case, the superposition of two plane-wave density fluctuations of equal amplitudes causes the overdensity at the intersection to be larger and, hence, the onset of nonlinearity to occur earlier than it would for either plane-wave alone. In the latter case, a *nonlinear* pancake perturbed by a *linear* amplitude symmetric mode described in §2.3.1 becomes *linearly* unstable with respect to fluctuation in its surface density. In this latter case, a dense clump eventually forms in the primary pancake central plane at the node of the transverse density perturbation long before the transverse perturbation has grown to the point of forming its own transverse pancake caustic. This indicates that dense clumps can grow not only at the intersections of pancakes of comparable size and collapse epoch as is widely believed, but also within individual pancakes, far from the nearest such intersection, as a result of a linear gravitational instability due to a much smaller transverse perturbation, long before the transverse perturbation itself grows nonlinear. This instability, moreover, occurs only for a limited range of unstable wavenumbers, a result anticipated by the energy argument, although the upper boundary of this range predicted by the energy argument proves to be incorrect as we shall see.

### 4.1.2. Antisymmetric Mode

We illustrate the qualitative outcome of antisymmetric mode perturbations by focusing on the case $k_a = 2$. Figures 5a and 5b show the particle positions at $a/a_c = 1$ and $a/a_c = 7$, respectively, for a pancake perturbed by mode $A_{2,0,2}$. In Figures 6a-e, we zoom-in on a subregion of the calculation box and display several time-slices spanning the range $1 \leq a/a_c \leq 7$. As shown in Figure 6, we find that, in this mode, the central primary pancake layer oscillates back and forth in the sense that it reverses its bending direction. In this case, the gravitational restoring force of the central pancake layer works to straighten the pancake, which results in an oscillation of the direction of particle trajectories. Meanwhile, by epoch $a/a_c = 4$, clumps form in the central pancake plane, at the maxima and minima of the y-variation of the potential $\phi$ as a result of the instability. We note that, in this mode, for unstable $k_a$, two clumps form per $\lambda_a$ (as opposed to only one per $\lambda_s$ for the symmetric mode).

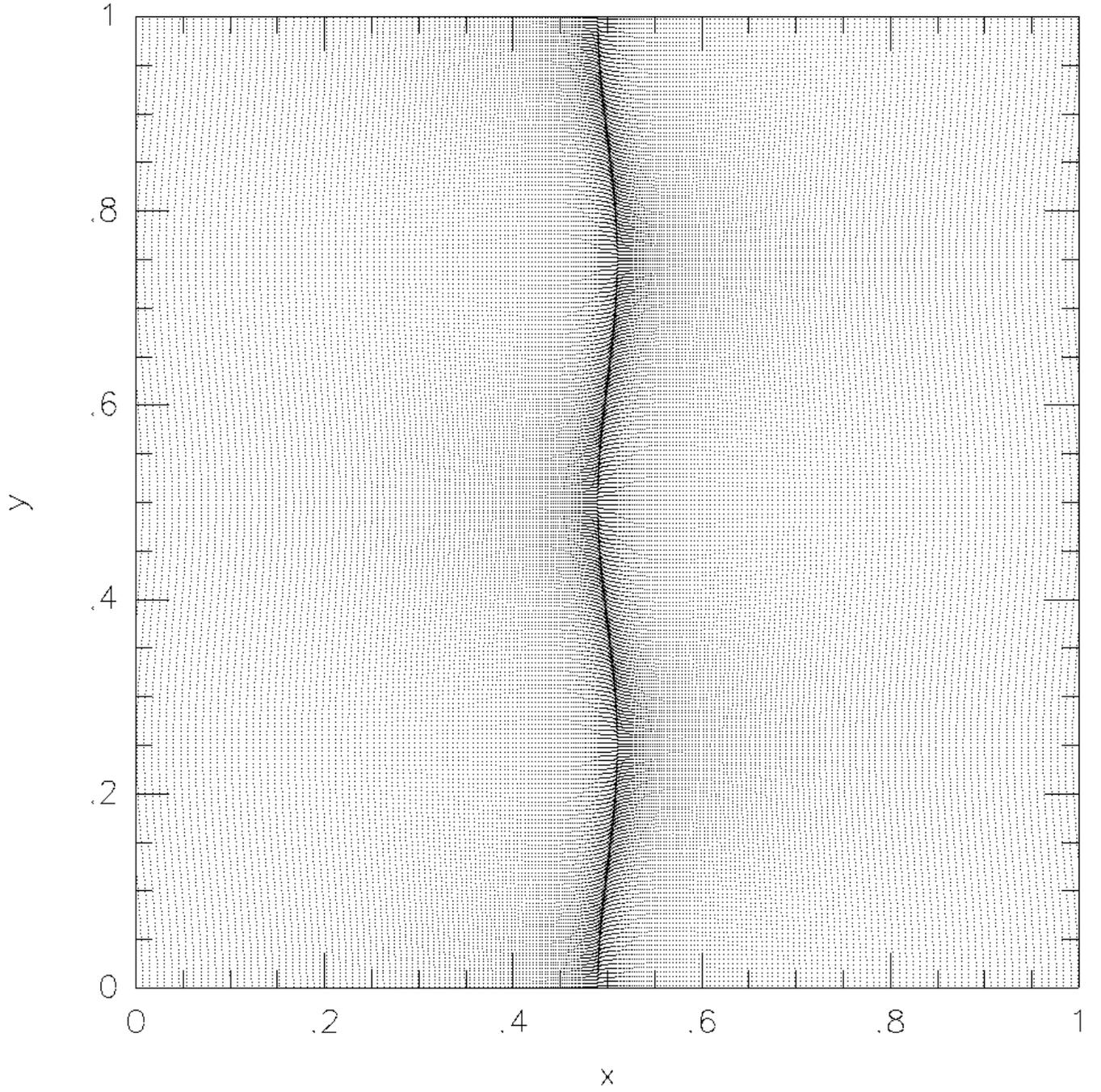

Fig. 5.— (a) (this page) Antisymmetric Mode. Particle positions at $a/a_c = 1$ for a PM simulation of a pancake perturbed by mode $A_{2,0.2}$. (b) (next page) Same as (a), but for $a/a_c = 7$.

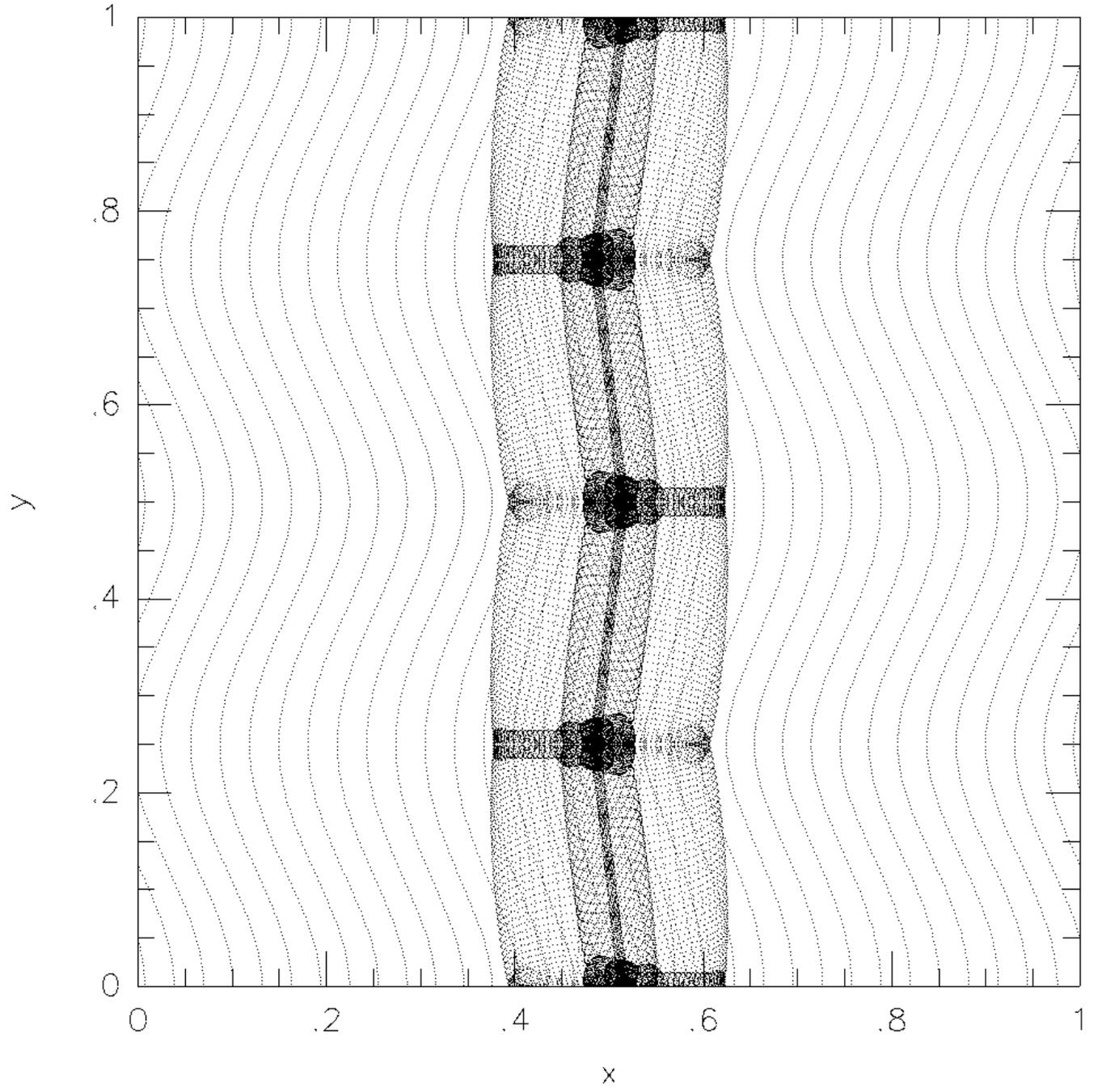

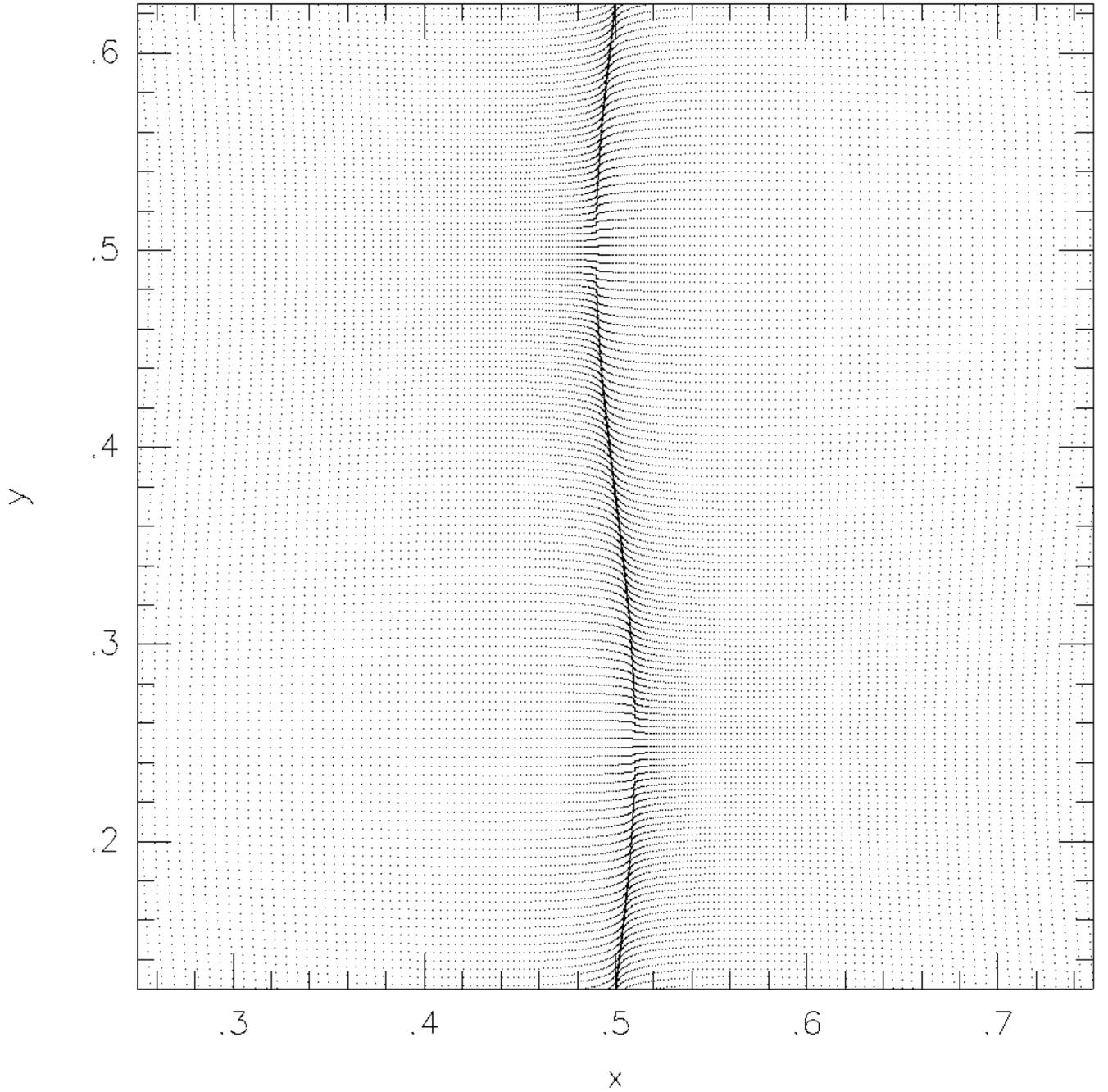

Fig. 6.— (a) (this page) Antisymmetric Mode Zoom-In. Particle positions at $a/a_c = 1$ for a PM simulation of a pancake perturbed by mode $A_{2,0.2}$, for a subregion of the computational box shown in Figure 5. (b), (c), (d), and (e) (next 4 pages) Same as (a), but for $a/a_c = 1.8$, $a/a_c = 2.5$, $a/a_c = 4.3$ and $a/a_c = 7.0$., respectively

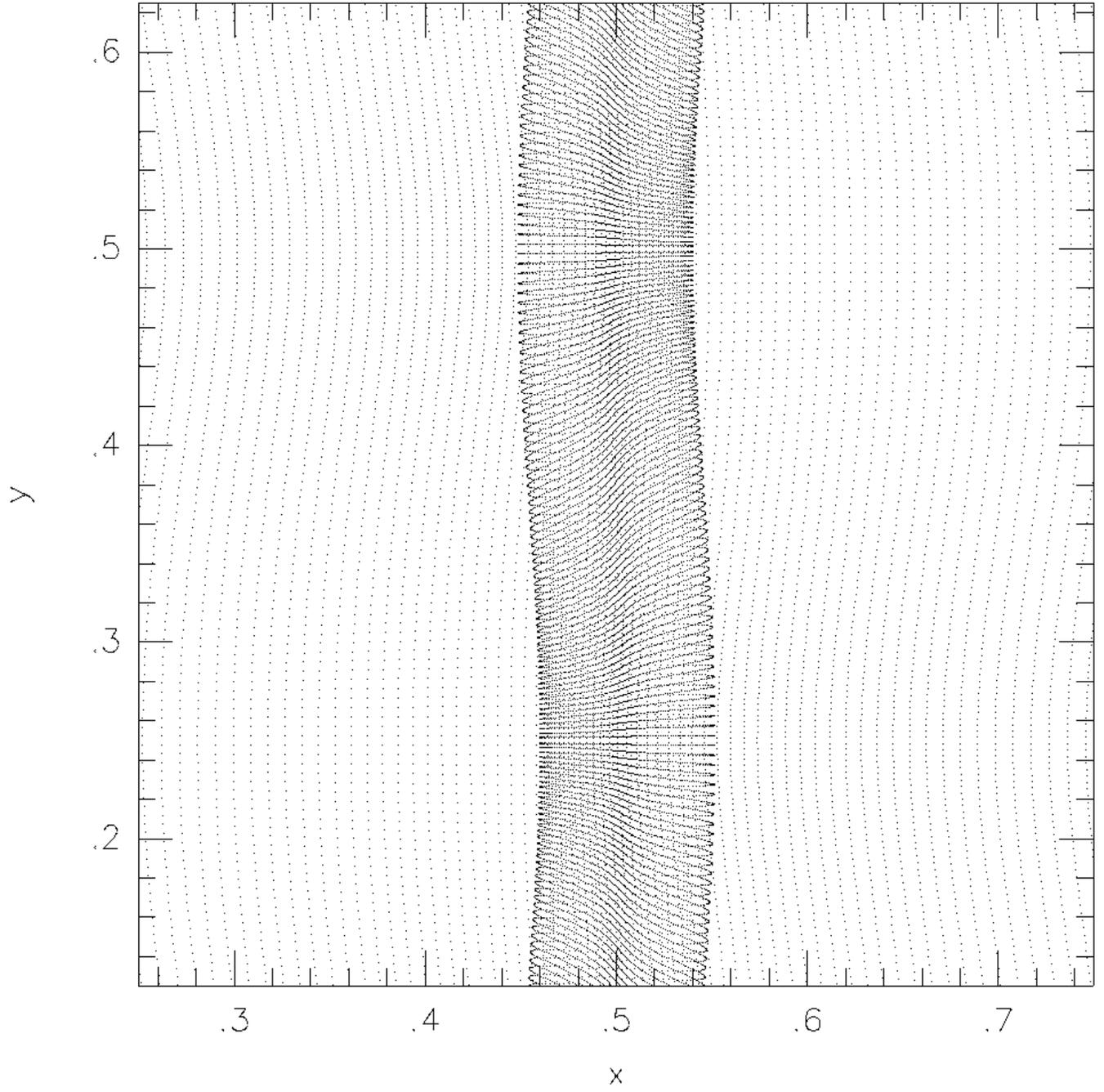

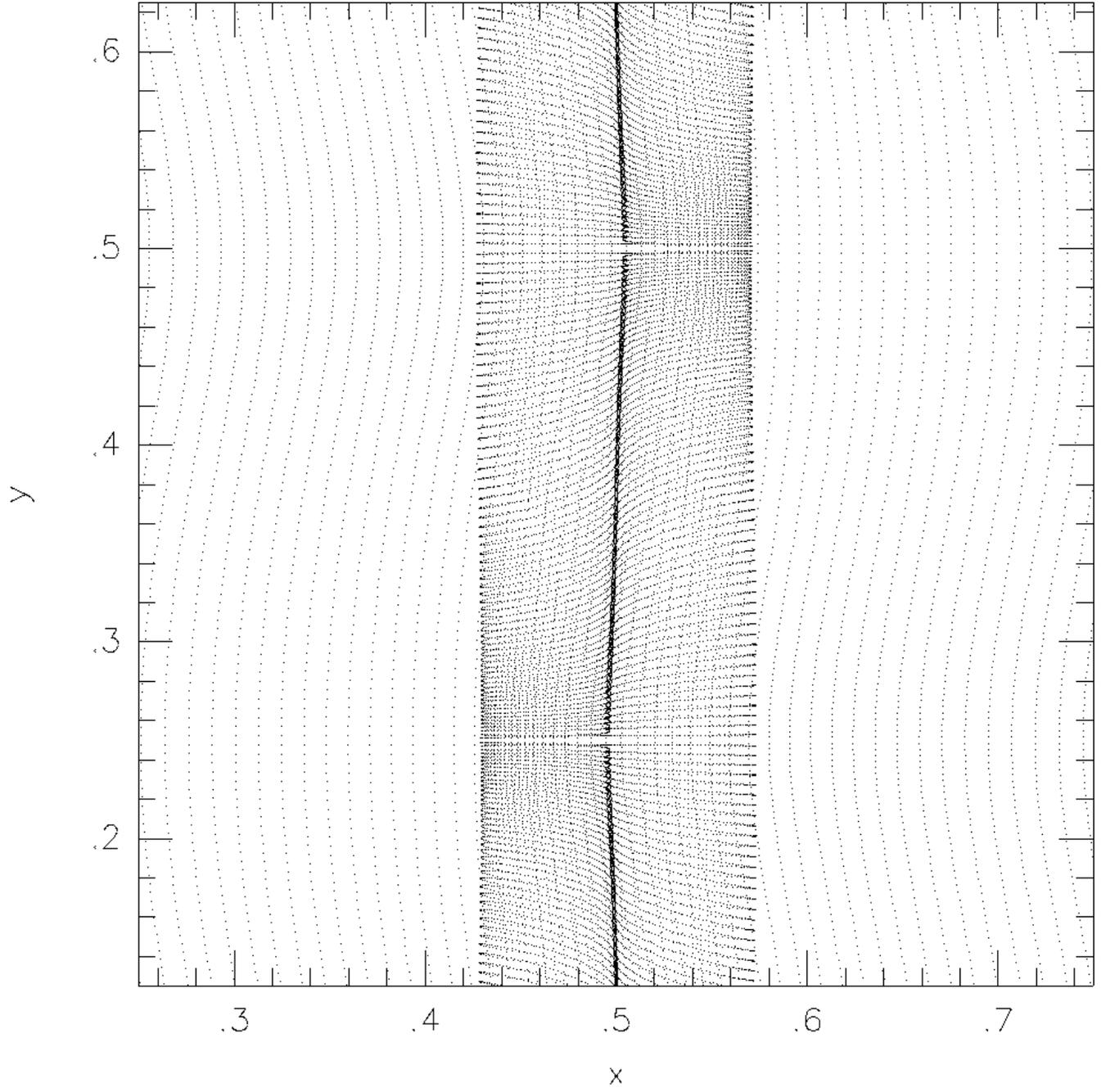

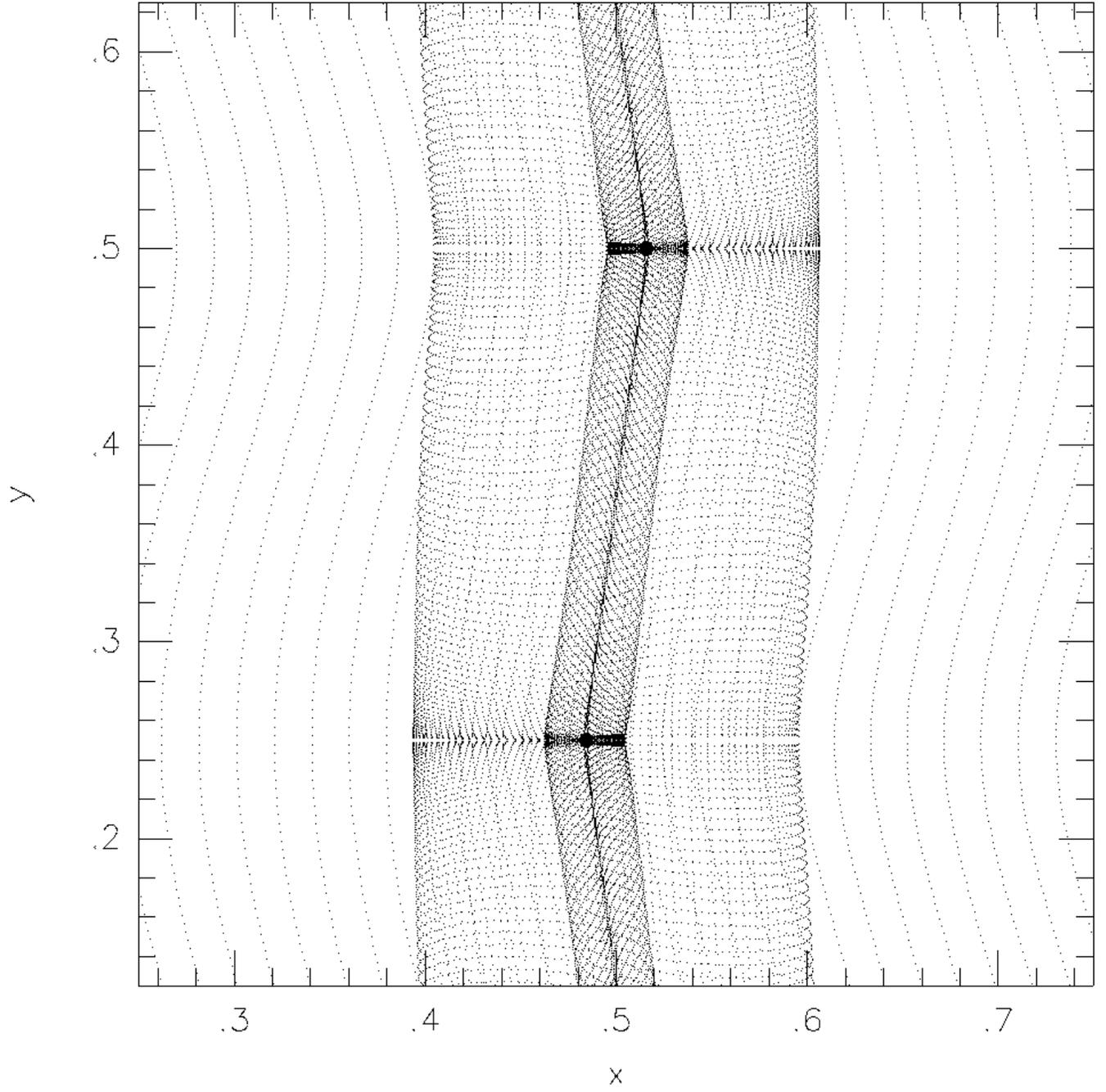

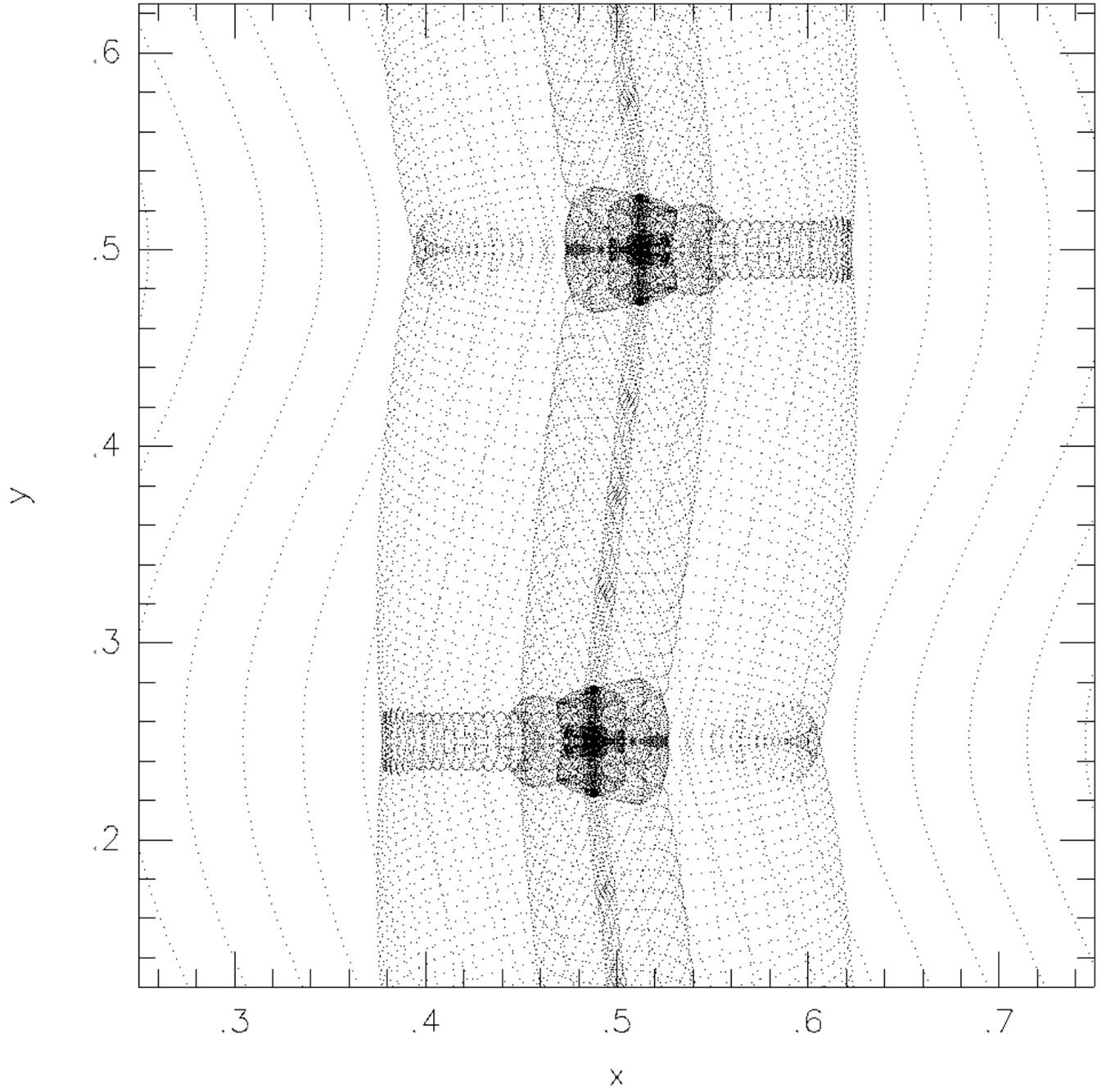

- 34 -## 4.2. Growth Rates and Epoch of Nonlinearity

### 4.2.1. Symmetric Mode

We focus first on the symmetric modes. The power spectrum analysis of the surface density fluctuation, $P_\Sigma$, indicates that for a given unstable $k_s$, the wavenumber of maximum power is $k = k_s$. This is evident from the plots of $P_\Sigma$ versus wavenumber $k$, at epoch $a/a_c = 2$ at which the instability is still in the linear regime, shown in Figure 7. In Figure 8, we have plotted the dimensionless growth rate, $\tilde{\Gamma}$, for $k = k_s$ (corresponding to the wavenumbers of maximum power for each $k_s$) vs. $a/a_c$ for a range of symmetric perturbation wavenumbers, $k_s = 1, 2, 4, 8, 16, 32,$ and $64$, where for each value of $k_s$, we display the results for five different amplitudes $\epsilon_s = 0.001, 0.01, 0.05, 0.1,$ and $0.2$. In general, the growth rate rises initially from the value of $\tilde{\Gamma} = 2/3$ expected for the linear growth rate of the transverse mode in the absence of the primary pancake. (Recall that, for a linear density fluctuation in the background universe, $|\delta\rho/\bar{\rho}|_{\max} = (a/a_c)$ and, therefore, $P(k) \propto (a/a_c)^2$, which leads to $\tilde{\Gamma} = (1/3)(a/a_c)[d\ln P(k)/d(a/a_c)] = 2/3$.) The growth rate rises until it reaches a maximum, flattens, and eventually declines (although for higher wavenumbers $k_s$ and high initial amplitude, the growth rate peaks and declines before it reaches the higher level of the plateau of the low-amplitude limit) and eventually declines. The epoch during which the growth rate rises signifies the onset of a new, *linear* instability. The flat part indicates the linear growth epoch. The epoch at which the dimensionless growth rate begins to decline marks the onset of nonlinear saturation of the instability growth. This epoch is coincident with the epoch at which $\sigma_n^2 = \sigma_\rho^2/\sigma_{\rho,0}^2 \geq 2$ as defined in §2.6. We have plotted $\sigma_n^2$ vs. $a/a_c$ for several perturbation modes of amplitude $\epsilon_s = 0.1$ in Figure 9 with the growth rate corresponding to each mode overlayed on the plot. It is evident that the decline in the growth rate corresponds to a sharp rise in the generated amount of nonlinearity. Hence, we shall call this epoch the onset of nonlinearity (i.e. $[a/a_c]_{NL}$). [6]

The variation of $\tilde{\Gamma}$ with initial amplitude $\epsilon_s$ for each given perturbation wavenumber

---

[6] The curves for the power spectra $P_\Sigma(k)$ versus time for a given perturbation wavenumber $k = k_s$ and its higher harmonics (not shown here) show that, as the instability approaches the point of nonlinear saturation, the decline of the growth rate for the wavenumber $k_s$ is accompanied by an increase in the growth rates for the higher harmonics, even though the power at $k = k_s$ still dominates. This transfer of power from long to short wavelengths reflects the steepening of the shape of the surface density perturbation with time. Such slowing of the growth rate of the fundamental perturbation mode and transfer of power from long to short wavelengths at the onset of nonlinearity is common to a wider range of problems involving gravitational instability. In a somewhat different context, similar behavior was reported for the case of the onset of nonlinearity during the formation of an unperturbed pancake (prior to caustic formation), by Shandarin & Zel'dovich (1989, p. 192), for example.

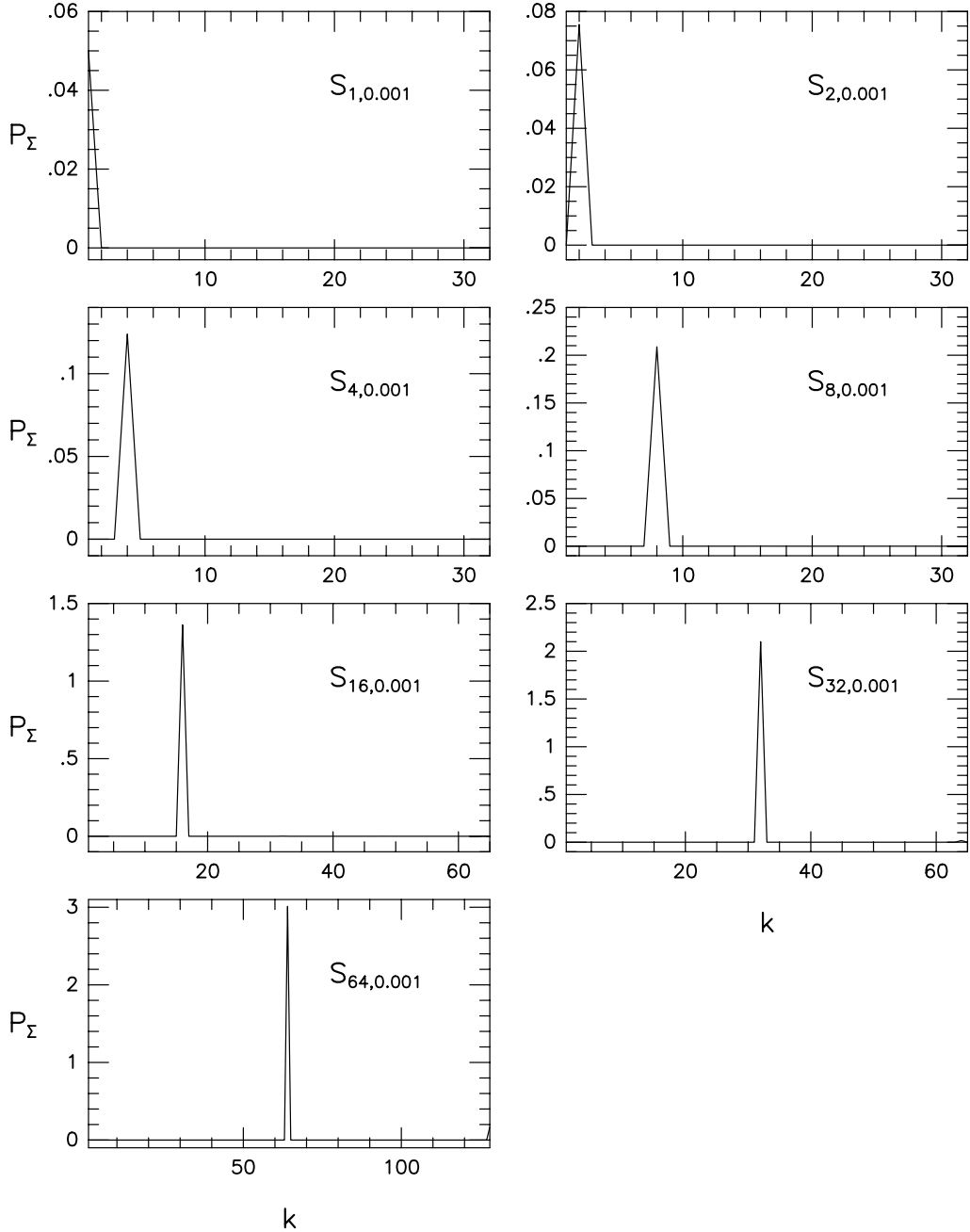

Fig. 7.— Power Spectrum $P_\Sigma(k)$ of the perturbation of pancake surface density is plotted against wavenumber $k$ for symmetric mode perturbations of fixed wavenumber $k_s$ for a single time-slice $a/a_c = 2.0$. Each panel represents a different perturbation mode $S_{k_s,\epsilon_s}$, as labelled, for $\epsilon_s = 0.001$ and $k_s = 1, 2, 4, 8, 16, 32$, and $64$, respectively.

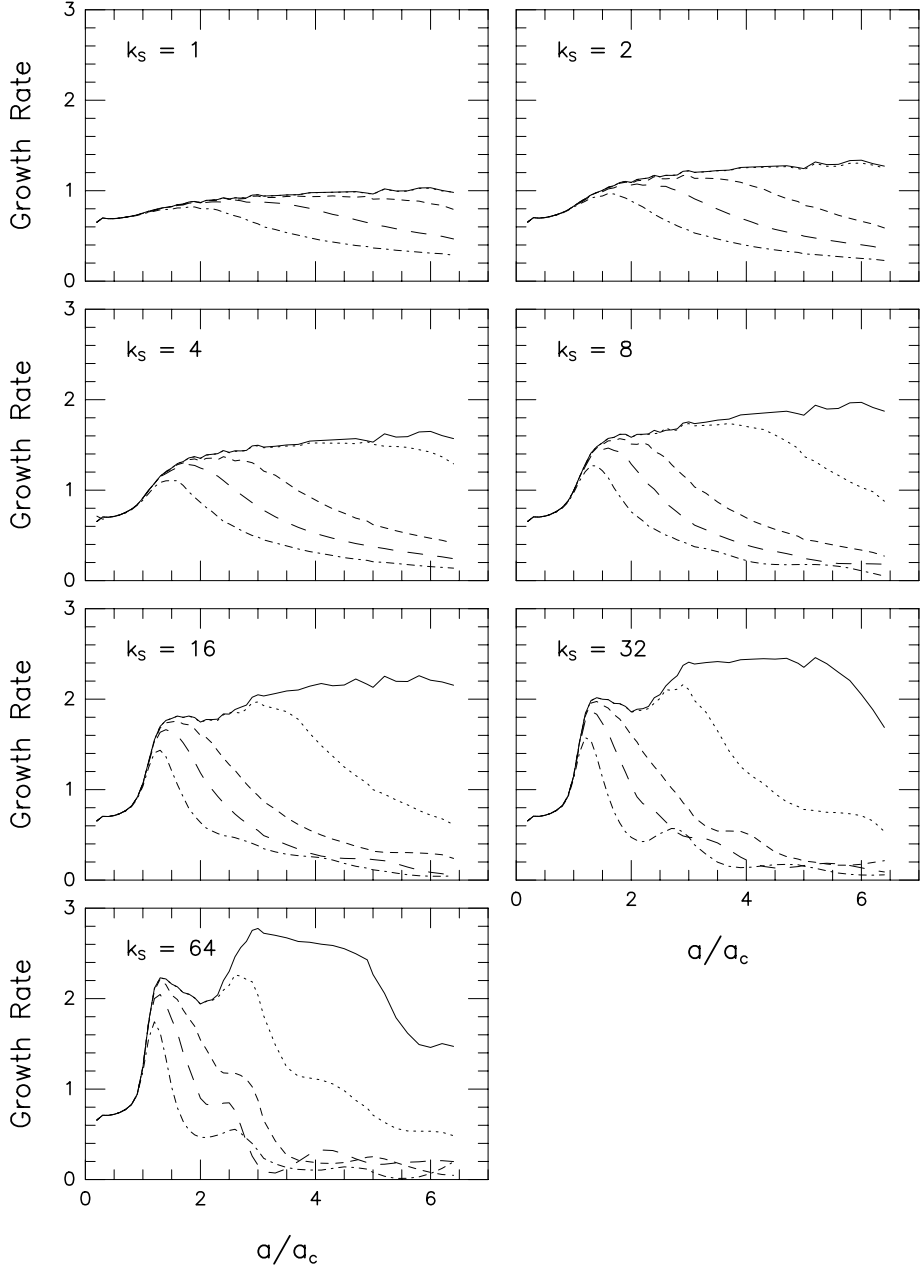

Fig. 8.— Dimensionless growth rate $\tilde{\Gamma}$ at wavenumber $k = k_s$, corresponding to the mode of maximum power for each symmetric mode perturbation wavenumber $k_s$, is plotted vs. $a/a_c$, for cases $S_{k_s,\epsilon_s}$, for $k_s = 1, 2, 4, 8, 16, 32$, and 64, respectively. Results are shown for initial amplitudes $\epsilon_s = 0.001$ (solid line), $\epsilon_s = 0.01$ (dotted line), $\epsilon_s = 0.02$ (short-dashed line), $\epsilon_s = 0.1$ (long-dashed line), and $\epsilon_s = 0.2$ (dot-dash line)

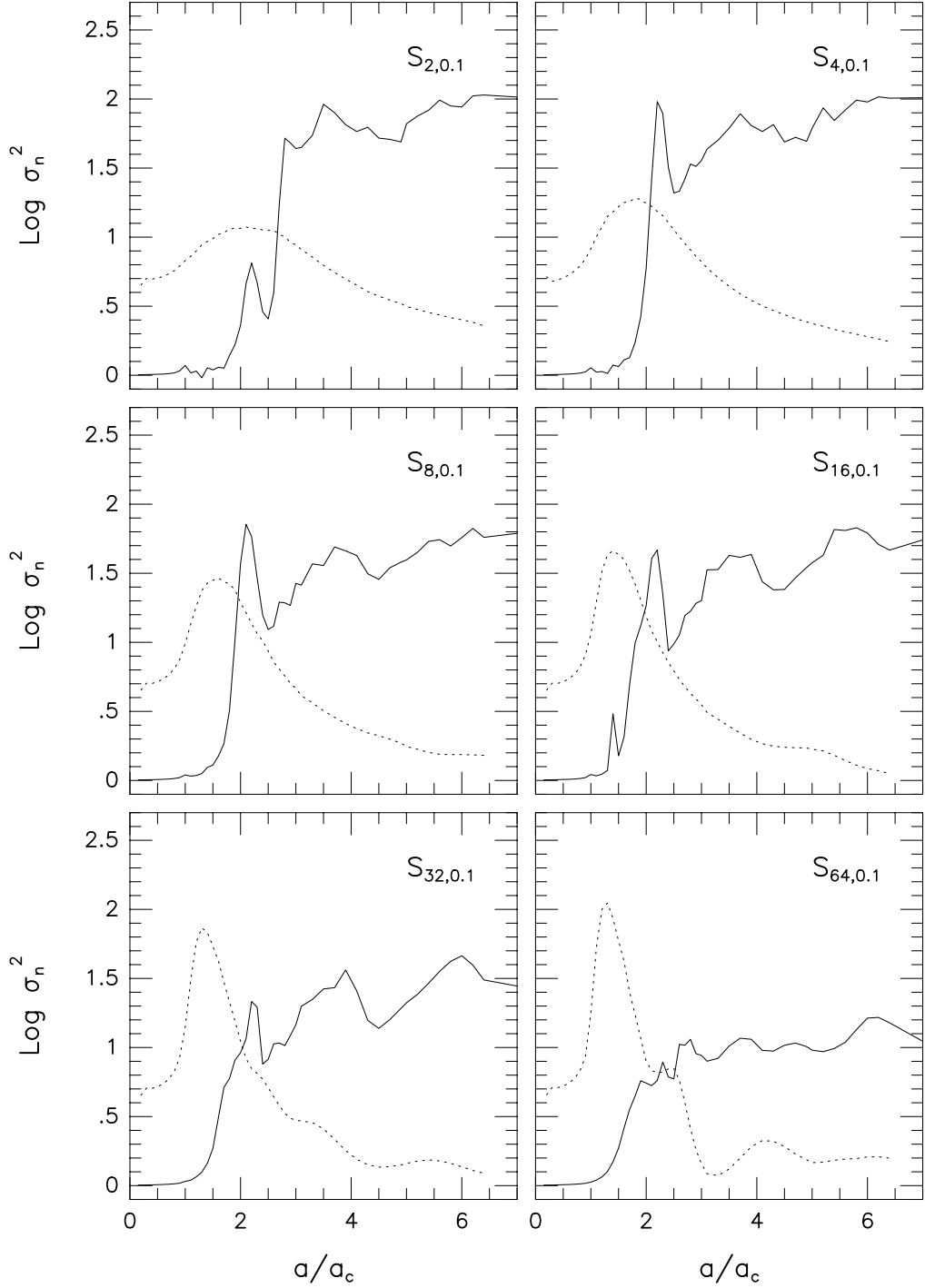

Fig. 9.— Normalized rms density fluctuation $\sigma_n^2$ is plotted vs. $a/a_c$ for symmetric modes as labelled (solid lines). Also plotted (dashed line) are the growth rates $\tilde{\Gamma}$ for $k = k_s$ for the same modes.

$k_s$, shown in Figure 8, is explained simply as follows. As the value of $\epsilon_s$ is decreased, the peak value of the dimensionless growth rate increases until, eventually, it converges to an asymptotic value which is initial-amplitude-independent and constant in time. This asymptotic value, the initial-amplitude-independent growth-rate "plateau," signifies that the instability is a true linear instability which has attained the full value of its growth-rate before it later declines due to the onset of nonlinear saturation. This asymptotic value of $\tilde{\Gamma} = \alpha =$ constant indicates that the fluctuation $\delta_\Sigma$ of the pancake surface density grows due to this instability as a power-law in time, $\delta_\Sigma \propto t^\alpha$ (where $t$ is proper time). As plotted in Figure 10, our results for this $\alpha$, the asymptotic, plateau-value of $\tilde{\Gamma}$ for the small $\epsilon_s$ limit, indicate that the dependence of $\alpha$ on perturbation wavenumber $k_s$ is a power law in $k_s$, which we have fit with the approximate formula $\alpha_s = \alpha_{s,1} k_s^n$, where $\alpha_{s,1} = 1.08$ and $n = 0.24$. (Incidentally, if we exclude the data point at largest wavenumber, i.e. $k_s = 64$, out of concern that there may not be sufficient resolution at small scales, the fit will yield $n \sim 0.25$). This is in contrast to $\tilde{\Gamma} \sim k^{1/2}$ as expected from the energy argument analysis (see eq. [47]).

### 4.2.2. Antisymmetric Mode

For unstable antisymmetric modes with a given perturbation wavenumber $k_a$, the wavenumber of maximum power is always $k = 2k_a$. This is evident from the plots in Figure 11 of the power spectra $P_\Sigma(k)$ versus $k$ at $a/a_c = 2.0$, for antisymmetric perturbation modes for different values of $k_a$. The dimensionless growth rate $\tilde{\Gamma}$ at the wavenumber $k = 2k_a$ is plotted in Figure 12 vs. $a/a_c$ for perturbation wavenumbers $k_a = 1, 2, 4, 8, 16, 32,$ and $64$ and, for each $k_a$, for initial amplitudes $\epsilon_a = 0.05, 0.1,$ and $0.2$ (and, for $k_a = 64$, $\epsilon_a = 0.01$, as well). Similar to the behavior described above for the symmetric modes, the growth rate rises initially from the value $\tilde{\Gamma} = 2/3$ (the value for the growth of a linear density fluctuation in the absence of the pancake mode), then levels off (or merely peaks, for large wavenumbers and higher initial amplitudes) and eventually declines. The rise signifies the onset of a new, *linear* instability. The levelling-off which occurs in the limit of small initial amplitude for each value of $k_a$ represents the attainment of the true linear growth rate (which is independent of initial amplitude). Our results for this asymptotic, amplitude-independent dimensionless growth rate $\tilde{\Gamma} = \alpha =$ constant indicate that unstable antisymmetric modes, too, grow as a power-law in time. As plotted in Figure 13, the power-law index $\alpha$ varies with perturbation wavenumber as a power-law in $k_a$, which we have fit with the relation $\alpha_a = \alpha_{a,1} k_a^n$, where $\alpha_{a,1} = 1.35$ and $n = 0.23$. (Incidentally, if we exclude the data point at largest wavenumber, i.e. $k_a = 64$, out of concern that there may not be sufficient resolution at small scales, the fit yields $n \sim 0.22$).

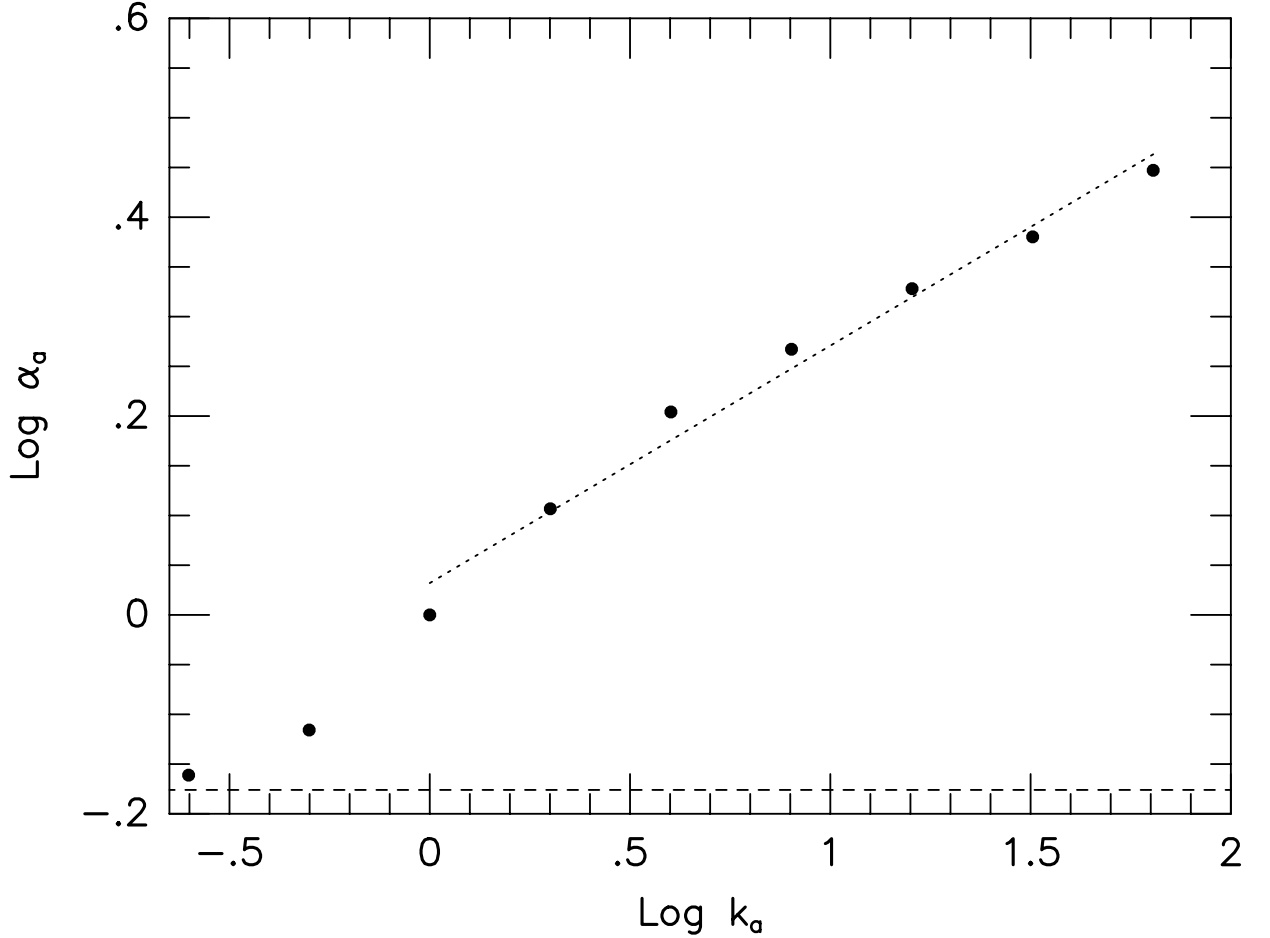

Fig. 10.— Asymptotic dimensionless growth rate $\tilde{\Gamma} = \alpha_s$ (i.e. amplitude-independent plateau-value for low amplitude limit) for $k = k_s$ is plotted vs. perturbation mode wavenumber $k_s$ for unstable symmetric modes with $1 \leq k_s \leq 64$. Simulation results (circles) are fit by the short-dashed straight line $\alpha_s = \alpha_{s,1} k_s^n$, where $\alpha_{s,1} = 1.08$ and $n = 0.24$. The horizontal long-dashed line at $\alpha_s = 2/3$ corresponds to the expected growth rate for density perturbations in the absence of the primary pancake mode.

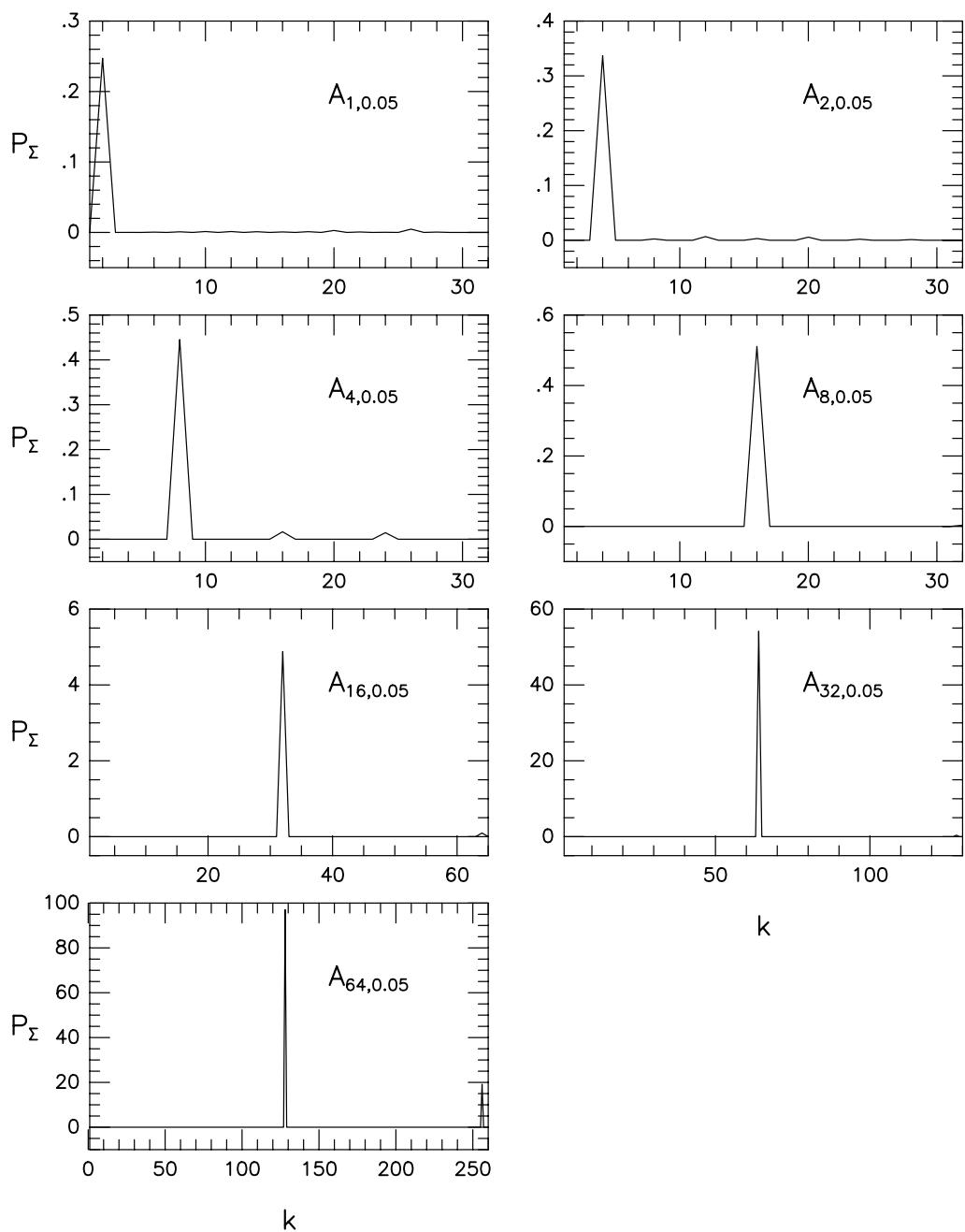

Fig. 11.— (a) Power Spectrum $P_\Sigma(k)$ of the perturbation of pancake surface density is plotted against wavenumber $k$ for antisymmetric mode perturbations of fixed wavenumber $k_a$ for a single time-slice. Each panel represents a different perturbation mode $A_{k_a,\epsilon_a}$, as labelled, for $\epsilon_a = 0.05$ and $k_a = 1, 2, 4, 8, 16, 32,$ and 64, respectively. The time-slice for $k_a = 1, 2, 4, 8,$ and 16 is $a/a_c = 2.0$ while that for $k_a = 32$ and 64 is $a/a_c = 2.5$.

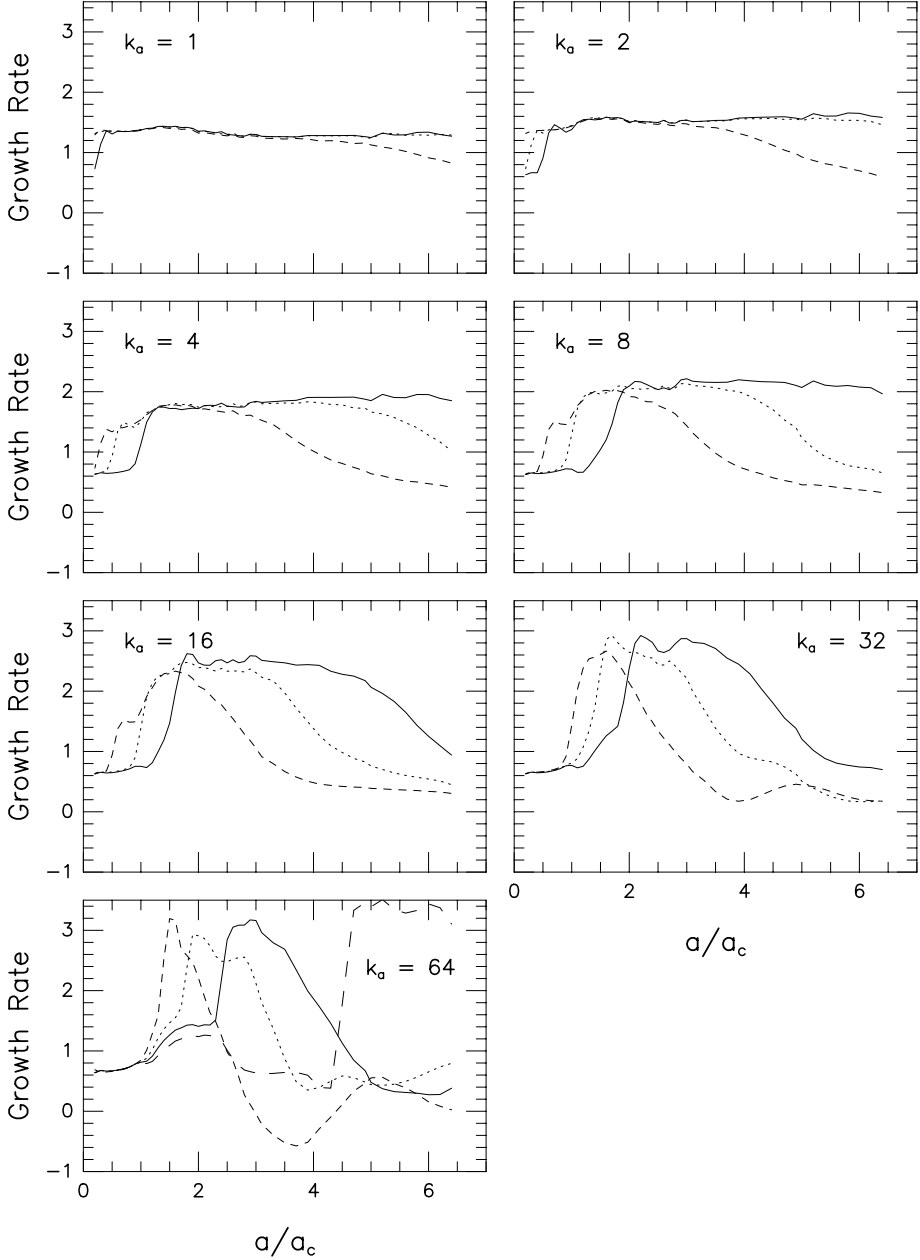

Fig. 12.— Dimensionless growth rate $\tilde{\Gamma}$ at wavenumber $k = 2k_a$, corresponding to the mode of maximum power for each antisymmetric mode perturbation wavenumber $k_a$, is plotted vs. $a/a_c$, for cases $A_{k_a,\epsilon_s}$, for $k_a = 1, 2, 4, 8, 16, 32$, and 64, respectively. Results are shown for initial amplitudes $\epsilon_a = 0.05$ (solid line), $\epsilon_a = 0.1$ (dotted line), and $\epsilon_a = 0.2$ (short-dashed line). For $k_a = 64$, an additional curve is shown for the case $\epsilon_s = 0.01$ (long-dashed line).

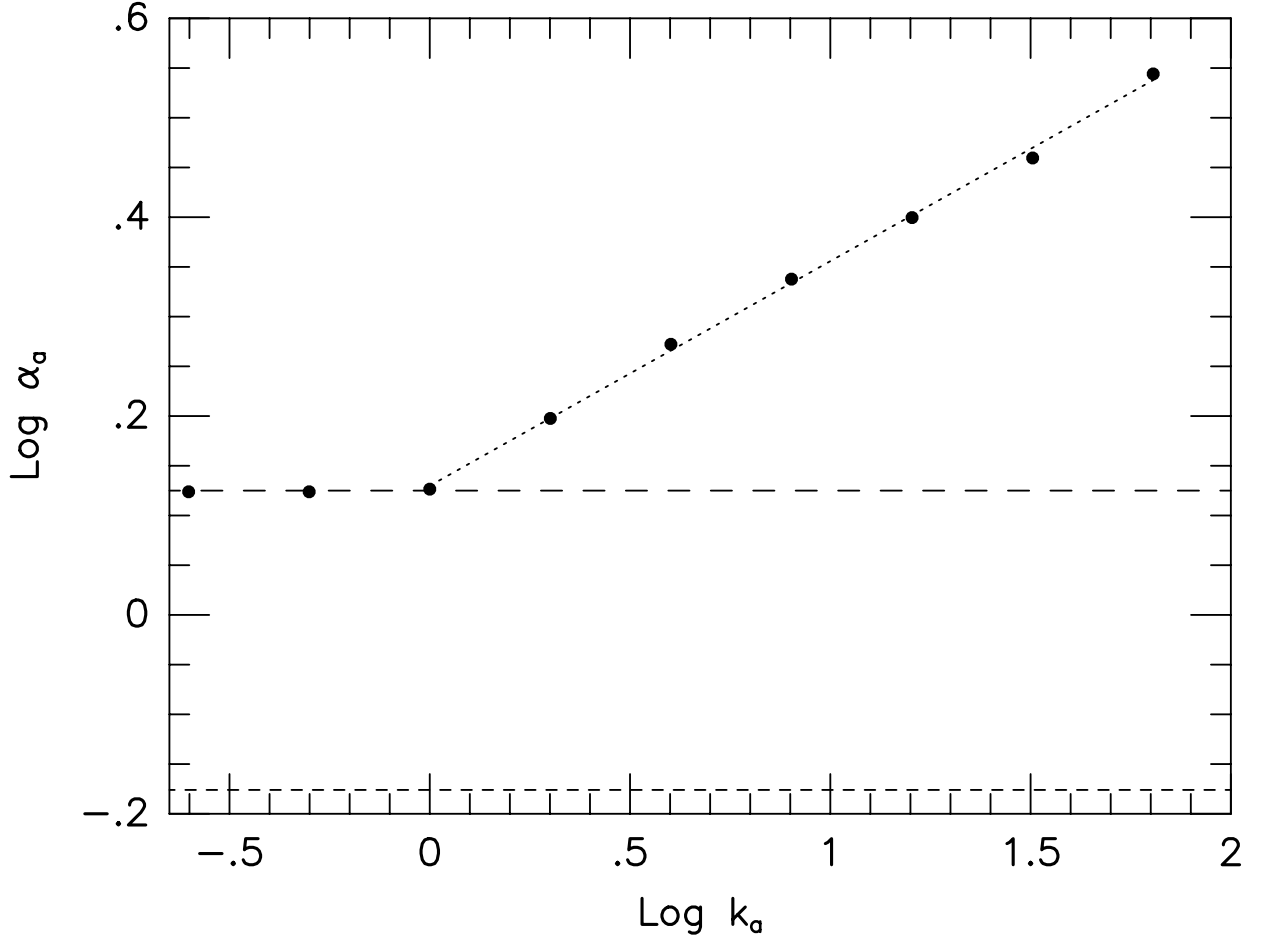

Fig. 13.— Same as Fig. 10, except for $k = 2k_a$ for the antisymmetric modes, plotted against $k_a$. The short-dashed line is the fit $\alpha_a = \alpha_{a,1} k_a^n$, where $\alpha_{a,1} = 1.35$ and $n = 0.23$. The horizontal long-dashed line at $\alpha_a = 4/3$ indicates the nominal growth rate due to the second-order (i.e. quadratic) term for the growth of density fluctuations for antisymmetric modes to which the pancake is not unstable.

For symmetric and antisymmetric modes of equal perturbation wavenumber and initial amplitude ($\epsilon_a = \epsilon_s$), the epoch of nonlinearity, $(a/a_c)_{NL}$, is reached later for the antisymmetric mode than for the symmetric mode. In other words, the linear growth phase – i.e. the flattened part of the growth rate plots – is much more extended for the antisymmetric mode than for the symmetric mode. This is due to the fact that the precaustic density perturbation for the symmetric mode is of first order in the perturbation amplitude while, it is only of second order for the antisymmetric mode (eqs. [26] and [39]), giving the symmetric mode an effectively larger initial density fluctuation.

### 4.2.3. Unstable Wavenumber Range

Pancakes in an initially "cold," collisionless gas are shown here to be gravitationally unstable when perturbed by either symmetric or antisymmetric transverse modes for a wide range of perturbation wavenumber. Our simulation results described above show that the range of unstable wavenumbers is not restricted on the high wavenumber side ($k \gg 1$). On the small wavenumber side, however, the instability is absent for perturbations with $k \lesssim 1$. We illustrate this in Figure 14. In the left panels, we have plotted the dimensionless growth rate $\tilde{\Gamma}$ versus $a/a_c$ for $k = k_s$ (or $k = 2k_a$ for the antisymmetric case) for symmetric mode cases $S_{1,0.05}$, $S_{1/2,0.05}$, $S_{1/4,0.05}$ and antisymmetric mode cases $A_{1,0.05}$, $A_{1/2,0.05}$, $A_{1/4,0.05}$. The normalized rms density fluctuation $\sigma_n^2$ is plotted against $a/a_c$ for these cases in the right panels.

As shown in Figure 14, the growth rates for the symmetric modes $k_s = 1/2$ and $k_s = 1/4$ are reduced, compared to that for $k_s = 1$. In fact, by $k_s = 1/2$, the growth rate is only marginally higher than the value of 2/3 for the growth rate of this mode in the *absence* of the primary pancake mode. A value of 2/3 for this growth rate signifies that pancakes are not unstable to perturbation by these modes.

The results for the antisymmetric cases also indicate that pancakes are not unstable when perturbed by modes with $k_a \lesssim 1$, but the interpretation of the plots in Figure 14 is more subtle than for the symmetric modes. In the antisymmetric cases, the linear perturbation in gravitational potential described in §2.3.2 is such that the surface density $\Sigma$ defined by equation (40) is unperturbed to linear order in the amplitude of the potential fluctuation. As a result, when a pancake is perturbed by a mode to which it is not unstable, the evolution of the power spectrum $P_\Sigma(k)$ and, hence, the dimensionless growth rate $\tilde{\Gamma}$ are initially dominated by the second-order growth rate for density fluctuations. As such, we expect the growth rate $\tilde{\Gamma}$ to be 4/3 (i.e. twice the value of 2/3 for the linear growth rate of density fluctuations in the absence of the pancake mode). This is precisely what the curves in Figure 14 (left panels) for $k_a \lesssim 1$ show. Eventually, the surface density fluctuation grows due to this quadratic term enough that the power spectrum $P_\Sigma(k)$ is sensitive to the first order growth rate, as well, and the value of $\tilde{\Gamma}$ declines so as to approach the value of 2/3. As shown in Figure 14, the normalized rms density fluctuation remains linear throughout the simulation for the antisymmetric modes $k_a = 1$, 1/2 and 1/4 (i.e. the departure of the rms density fluctuation from that of an unperturbed pancake is always very small). Hence, the gradual decline of $\tilde{\Gamma}$ from 4/3 (the nominal value for the second-order growth rate expected if the pancake is *not* unstable to these perturbation modes) to the value of 2/3 (the nominal value for the linear growth rate in that case) which appears in Figure 14 for $k_a = 1/2$ and 1/4 is *not* a reflection of nonlinear saturation. Lastly, we note that the value of $\tilde{\Gamma} = 4/3$ is achieved by the antisymmetric modes at epochs significantly in advance of the epoch of pancake caustic formation, so this value of 4/3 is clearly not a response to the nonlinear evolution of the pancake mode. In short, the results in Figure 14 indicate that pancakes are not unstable to perturbation by antisymmetric modes with $k_a \lesssim 1$.

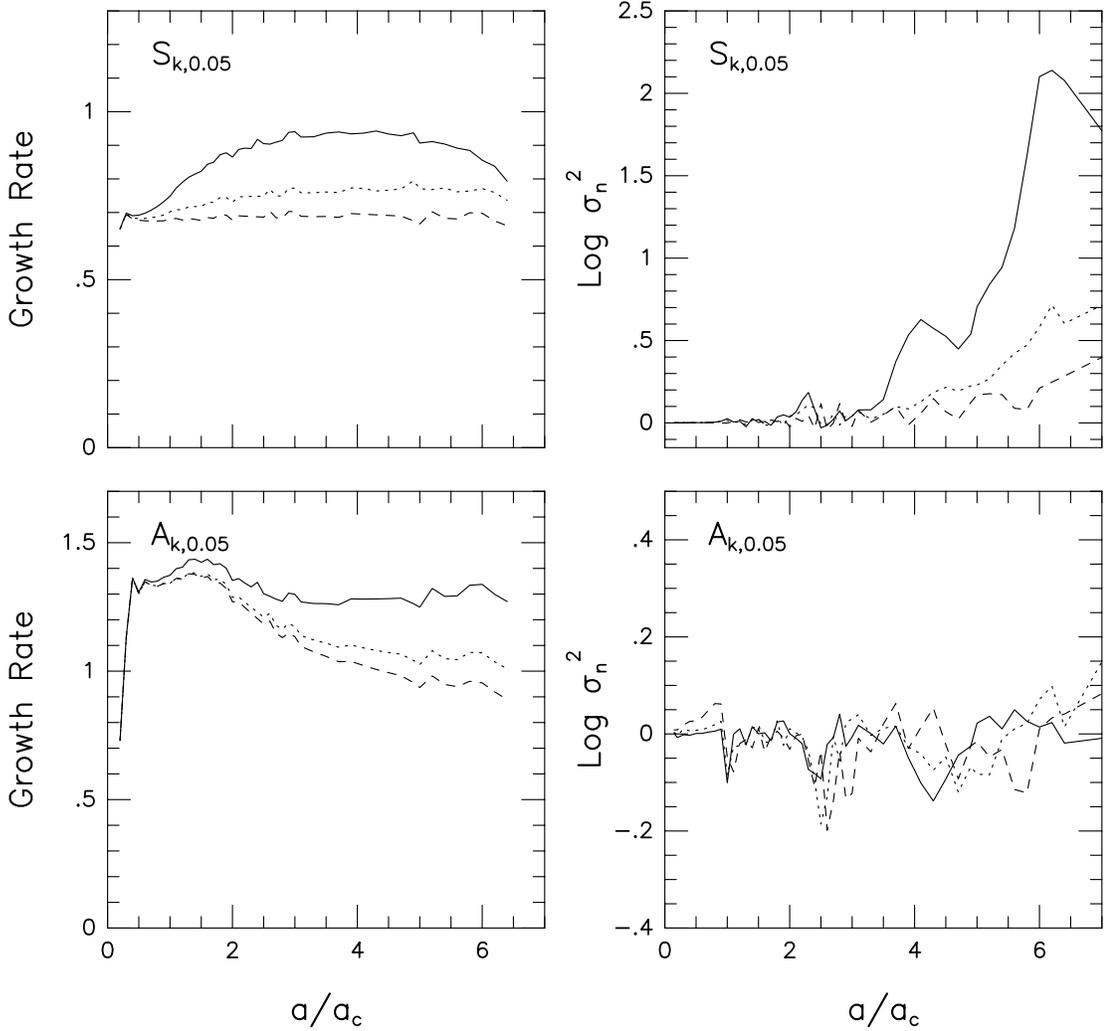

Fig. 14.— Left panels: Dimensionless growth rate $\tilde{\Gamma}$ vs. $a/a_c$ for symmetric modes at $k = k_s$ (upper left panel) and antisymmetric modes at $k = 2k_a$ (lower left panel). In the upper left panel, we plot cases $S_{1,0.05}$ (solid curve), $S_{1/2,0.05}$ (dotted curve) and $S_{1/4,0.05}$ (dashed curve). In the lower left panel, we plot cases $A_{1,0.05}$ (solid curve), $A_{1/2,0.05}$ (dotted curve), and $A_{1/4,0.05}$ (dashed curve). Right panels: Normalized rms density fluctuation $\sigma_n^2$ vs. $a/a_c$ for the same cases as plotted in the left panels. Upper right panel (symmetric) and lower right panel (antisymmetric) use the same line types to indicate the different cases as in the left panels.

### 4.3. Is the Thin-Sheet Energy Argument Adequate?

We now compare our simulation results for the "cold" limit (i.e. collisionless particles with zero initial, random thermal velocity) with the predictions of the thin-sheet energy argument. As shown in §3 and Figure 2, the thin-sheet energy argument predicts that the pancake is gravitationally unstable for $k_{\min} < k < k_{\max}$. When the argument was applied to the results of 1D, unperturbed pancake simulations in §3.2, the value of $k_{min}$ was found to be of order unity (i.e. $k_{min} \approx 0.6$). Our 2D simulation results and the analysis described above show rough agreement with this prediction on the small wavenumber side, that modes with perturbation wavevector $k < k_{min} \sim 1$ are not unstable. We also find that modes with $k > k_{min}$ are indeed unstable. However, the simulations presented here show that modes of wavenumber $k > k_{\max}$ are also unstable. The positive energy contributed by the 1D velocity dispersion of the unperturbed pancake matter, moving along the direction of pancake collapse within the region of shell-crossing, is apparently not enough to prevent the pancake from being gravitationally unstable, contrary to the expectations of the energy argument.

This latter result is similar to that found by Hwang et al (1989) in their linear modal analysis of gravitational instability in a self-similar, spherical, expanding, cosmological void in collisionless matter. That work showed that 1D velocity dispersion in the radial direction within the region of shell crossing could not prevent the focusing of particle trajectories, despite the expectations of the energy argument, resulting in the growth of nonradial per-turbation. In that case, the dimensionless growth rate of the perturbation was found to vary with the spherical eigennumber $l$ (the spherical equivalent of wavenumber) as $\ln^{1/2}(1.78/l)$. We find for the pancake problem that the growth rate scales with wavenumber as $k^n$, where $n \approx 0.2 - 0.25$ for both symmetric and antisymmetric modes. Hence, the increase of growth rate with decrease of perturbation scale length is faster for a pancake than for a spherical shell.

For all modes which are unstable, we also find that our numerical result for the dependence of linear growth rate on perturbation wavenumber $k$ departs significantly from that predicted by the thin-sheet energy argument. The energy argument predicts that the dimensionless growth rate scales roughly as $k^{1/2}$, while our simulations find that a scaling as $k^n$ with $n \approx 0.2 - 0.25$ is a more accurate description.

In short, the thin-sheet energy argument failed to anticipate that collisionless pancakes are unstable to perturbations with $k > k_{\max}$ when the matter from which they form is "cold". This fact, together with the earlier, analogous finding by Hwang et al. (1989), suggested to us that the purely 1D velocity dispersion of counter-streaming matter within the region of caustics and shell-crossing in the pancake might play a role which is fundamentally different from that of the isotropic, random motions associated with a finite initial temperature for the matter. The latter, thermal motions, we conjectured, might act to stabilize the pancake against transverse perturbation in closer agreement with the energy argument, even though counter-streaming 1D motions of initially cold matter did not. We address this possibility in the following section.

## 4.4. Effect of Initial Thermal Velocity

Let us now discuss briefly the effect of a non-zero initial temperature on the outcome of the gravitational instability for the modes described in this paper. To investigate this effect, we have performed a series of numerical simulations in which we added a thermal velocity component to the initial streaming velocity of the collisionless particles. In order to represent the spread in magnitude and direction of the random, isotropic, thermal velocity distribution at each location in this case, we initialized our simulation particles as follows. Six particles of equal mass were initially placed at the centers of each cell of a uniform square lattice. Each set of six particles was perturbed in location and velocity so as to represent the pancake and transverse modes as described in §2. We then divided the six particles into three pairs. For each pair, equal and opposite additional velocities were added by randomly sampling an isotropic Maxwellian distribution of thermal velocity by a Monte Carlo program. A spatially uniform temperature was assumed for this Maxwellian distribution.

In order to parametrize the relative amount of thermal velocity assumed, we have defined an effective "Mach number" $M$ according to

$$M = \frac{\langle v_s^2 \rangle^{1/2}}{\langle v_{\text{th}}^2 \rangle^{1/2}},\tag{54}$$

where $v_s$ is the streaming velocity along the direction of collapse for an unperturbed pancake, $v_{\text{th}}$ is the thermal velocity, and the $\langle \rangle$ refers to an average over all of space. We shall henceforth refer to values of $M$ evaluated at $a/a_c = 1$ when using this parameter to describe the amount of random thermal velocity introduced as an initial condition in pancake calculations with finite initial temperature.

According to the thin-sheet energy argument in §3, the gravitational instability of a pancake should be suppressed if the initial temperature of the collisionless particles which formed the pancake is high enough. The minimum value of $v_{\text{rms}}$ which is required to suppress instability altogether is that which makes $k_{\min} = k_{\max}$. According to equations (48)-(51), this occurs when

$$v_{\text{rms}} = v_{\text{rms,crit}} = \frac{2^{1/2}\pi G K \Sigma}{H}.\tag{55}$$

The surface density of the central layer of the unperturbed pancake (i.e. the region of shell-crossing) can be written as

$$\Sigma = \rho \lambda_p f\left(\frac{a}{a_c}\right),\tag{56}$$

where $f(a/a_c)$ is the mass fraction within this layer at epoch $a/a_c$, where $f(a=a_c) = 0$ and $0 < f(a/a_c) < 1$ for $a/a_c > 1$. In that case, it can be shown,

$$\frac{v_{\text{rms,crit}}}{v_{H,p}} = \frac{3 \cdot 2^{1/2}}{8} f\left(\frac{a}{a_c}\right),\tag{57}$$

where $v_{H,p} \equiv H(t)\lambda_p(t)$.

Let us consider the "cold" limit (i.e. $M = \infty$) first. In that case, $v_{rms} = \langle v_s^2 \rangle^{1/2}$ and the results for 1D pancake collapse indicate that $\langle v_s^2 \rangle^{1/2} \sim v_{H,p}$ is typical for $a/a_c \gtrsim 1$ (see, for example, Fig. 1). According to equation (57), this means that the vertical motions induced by the pancake collapse should be roughly sufficient to suppress the instability for most perturbation wavenumbers at some time $a/a_c > 1$. This point is made more precise by the plots of $k_{\min}$ and $k_{\max}$ in Figure 2, based on the detailed results of 1D pancake simulations. According to Figure 2, pancakes should be stable for perturbation wavenumber $k > k_{\max}$, where $k_{\max}$ reflects the positive contribution to the thin sheet energy from these vertical motions induced by pancake collapse, and $k_{\max}$ is not much bigger than $k_{\min}$. It is remarkable, therefore, that this 1D vertical velocity dispersion proves *not* to be capable of suppressing the instability *at all*, according to our 2D simulation results. Our simulation results may indicate that the purely 1D nature of the velocity dispersion induced by pancake collapse and counter-streaming is not well represented by the thin-sheet energy argument. On the other hand, such an interpretation still leaves open the possibility that the contribution to $v_{\rm rms}$ from isotropic, random thermal motions, $\langle v_{\rm th}^2 \rangle^{1/2}$, can act to suppress the instability in the manner expected from the energy argument even if the 1D contribution to $v_{\rm rms}$ from streaming motions cannot.

Let us consider the effect of finite temperature (i.e. $M < \infty$), therefore. According to equations (54) and (57), we expect that, for $M \sim 1$, temperature stabilizes the pancake against gravitational fragmentation, since $\langle v_s^2 \rangle^{1/2} \sim v_{H,p}$. In this case, however, the pancake, itself, barely manages to form, since this high finite temperature implies that the effective Jeans length in the background, unperturbed universe, $\lambda_J$, is comparable to $\lambda_p$. For $M < 1$, in fact, we do not expect the pancake to form at all, since $\lambda_p < \lambda_J$ in that case. For $M \gtrsim 1$, however, we expect that the pancake will form and be gravitationally unstable, but only for modes in the range $k_{\min} < k < k_{\max}$, where $k_{\min}$ and $k_{\max}$ are given by equations (49)–(51), with $v_{\rm rms}$ replaced by $\langle v_{\rm th}^2 \rangle^{1/2}$. How does $k_{\max}$ depend upon the Mach number $M$? In the limit in which $k_{\max}/k_{\min} \gg 1$, it can be shown,

$$k_{\max} \cong \frac{\pi G K \Sigma \lambda_p}{v_{\rm rms}^2} = \frac{3}{8} K f\left(\frac{a}{a_c}\right) \frac{v_{H,p}^2}{v_{\rm th}^2}. \tag{58}$$

According to equations (54) and (58), therefore, the fact that $\langle v_s^2 \rangle^{1/2} \sim v_{H,p}$ for $a/a_c \gtrsim 1$ implies that we can write this as

$$k_{\max} \sim \frac{3}{8} K f\left(\frac{a}{a_c}\right) M^2. \tag{59}$$

Since the factor $3Kf(a/a_c)/8 < 1$, equation (59) implies that $k_{\max} \lesssim M^2$. In short, we expect from the thin-sheet energy argument that pancakes perturbed by modes with $k \gtrsim M^2 > 1$ will be stable against fragmentation.

In order to test this expectation, we have performed the following set of 2D PM simulations. We have simulated the response of pancakes in a collisionless gas of finite initial temperature to transverse perturbation by either symmetric or antisymmetric modes, for a fixed perturbation amplitude $\epsilon = 0.2$ at initial epoch $a_i/a_c = 0.15$, for temperatures corresponding to $M = 17, 6$, and $1.7$ (only $M = 17$ and $M = 6$ for the antisymmetric modes), and for $k = 1, 2, 4, 8, 16$, and $32$. For each simulation, we have calculated the dimensionless growth rate $\tilde{\Gamma}$ of the perturbation modes (i.e. for $k = k_s$ or $k = 2k_a$, respectively) versus time and plotted it together with the corresponding growth rate for the same pancake perturbation when no thermal velocity was present. Our results are shown in Figures 15 and 16 for the symmetric and antisymmetric modes, respectively. We also plot the power spectrum $P_\Sigma(k)$ versus wavenumber at a particular epoch, $a/a_c = 2$, in Figure 17, for the symmetric case only, for $k_s = 1, 2, 4$, and 8.

We focus on the symmetric mode results in our analysis. Our results are summarized as follows: (1) For pancakes of high enough Mach number (i.e. low enough temperature), the instability develops just as in the zero temperature limit for which detailed results were discussed earlier. The meaning of the term "high enough Mach number," however, is dependent on the wavenumber of the perturbation, increasing with wavenumber. For $k_s = 1$, for example, the perturbation growth is unaffected by finite temperature in our simulation results for $M \gtrsim 6$, while for $k_s = 8$, this is true for the cases simulated with $M \gtrsim 17$. (2) For $M < 1$, the temperature is so high that the primary pancake is prevented from forming, as expected. (3) For Mach numbers of intermediate value, however, where the energy argument predicts that instability is suppressed by thermal velocity if $k > k_{\max} \sim M^2$, we clearly find to our surprise that the pancakes are unstable just as in the zero temperature limit, and with the same linear growth rate as in that limit. This surprising result, which further contradicts the thin-sheet energy argument, is apparent only after the following details are made clear.

First, for perturbations with $k \gtrsim k_{\max} \sim M^2$, the perturbation modes decay prior to the formation of the primary pancake (i.e. between $a/a_c = a_i/a_c < 1$ and $a/a_c = 1$), since the transverse perturbation wavelengths $\lambda$ are such that $\lambda < \lambda_J$, the Jeans length in the unperturbed background universe. As a result, these perturbation modes have very much smaller amplitudes at $a/a_c = 1$ with which to perturb the pancake, compared with their amplitudes at $a/a_c = 1$ in the zero temperature limit. Nevertheless, as the curves for $M = 1.7$ and $k_s = 4$ and 8 in Figure 15 show most clearly, this reduced amplitude perturbation at the input perturbation wavenumber $k_s$, which is still present at epochs $a/a_c \geq 1$ eventually grows with the same dimensionless growth rate $\tilde{\Gamma}$ as found for the zero-temperature results. The differences amongst the growth rate curves for different values of $M$ but the same value of $k_s$, in the panels of Figure 15 for $k_s = 4$ and 8 are explained entirely by the fact that smaller values of $M$ correspond to perturbations of smaller amplitude at $a/a_c = 1$. The curves for $M = 1.7$ for $k_s = 4$ and 8, in fact, are higher at $a/a_c > 1$ than the others for larger $M$ values precisely because the amplitudes of the transverse perturbation modes at $a/a_c = 1$ for $M = 1.7$ were so small. The reduced-amplitude perturbations at $a/a_c = 1$ in this case place the instability well within the asymptotic, small-amplitude limit in which the growth rate rises to a maximum value which is amplitude-independent, and, thereafter levels off, reflecting the true linear growth rate.

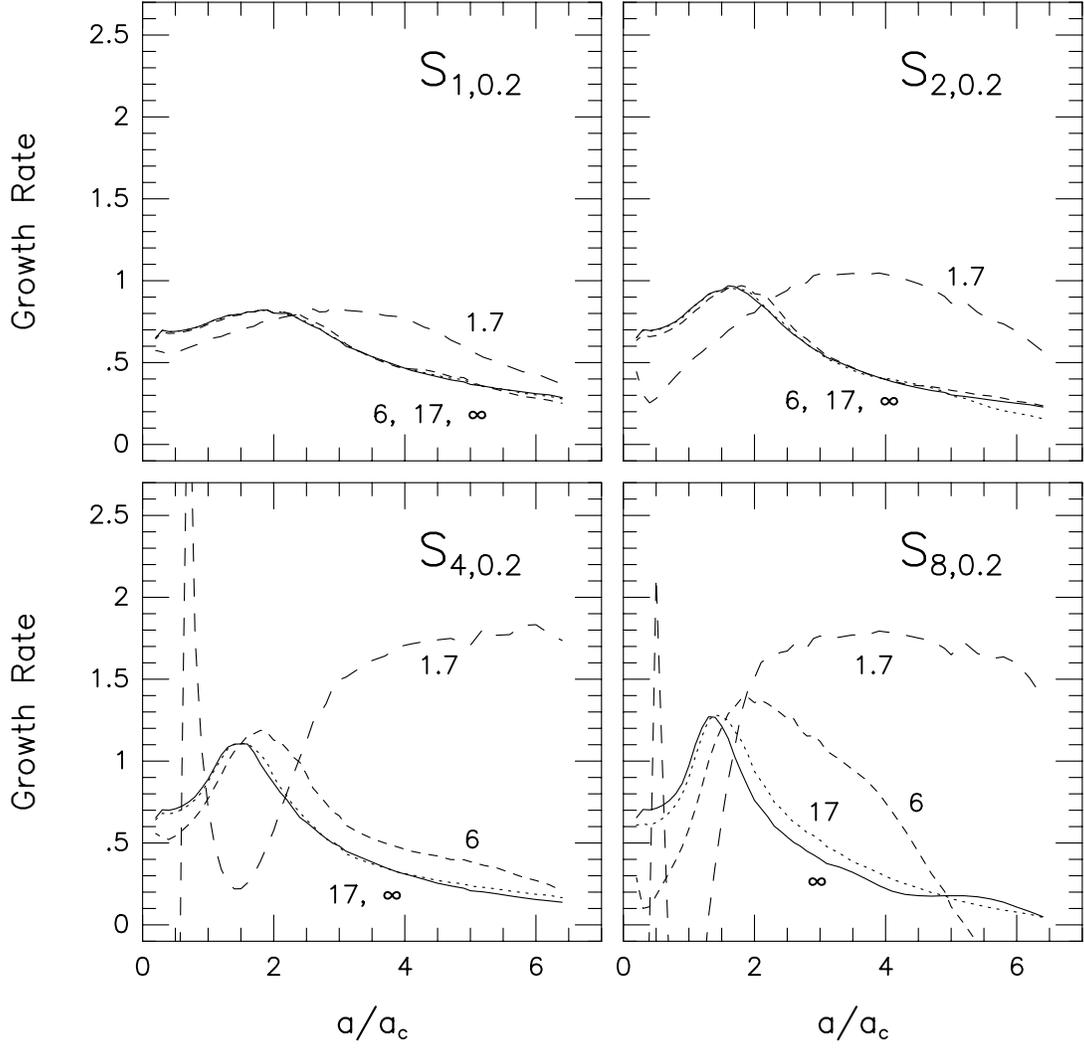

Fig. 15.— Dimensionless growth rate $\tilde{\Gamma}$ at wavenumber $k = k_s$ vs. $a/a_c$, for symmetric modes with finite temperature, for $M = \infty$ (i.e. no thermal velocity) (solid curve), $M \approx 17$ (dotted), $M \approx 6$ (short-dashed), and $M \approx 1.7$ (long-dashed), for modes $S_{1,0.2}$, $S_{2,0.2}$, $S_{4,0.2}$, and $S_{8,0.2}$, as labelled.

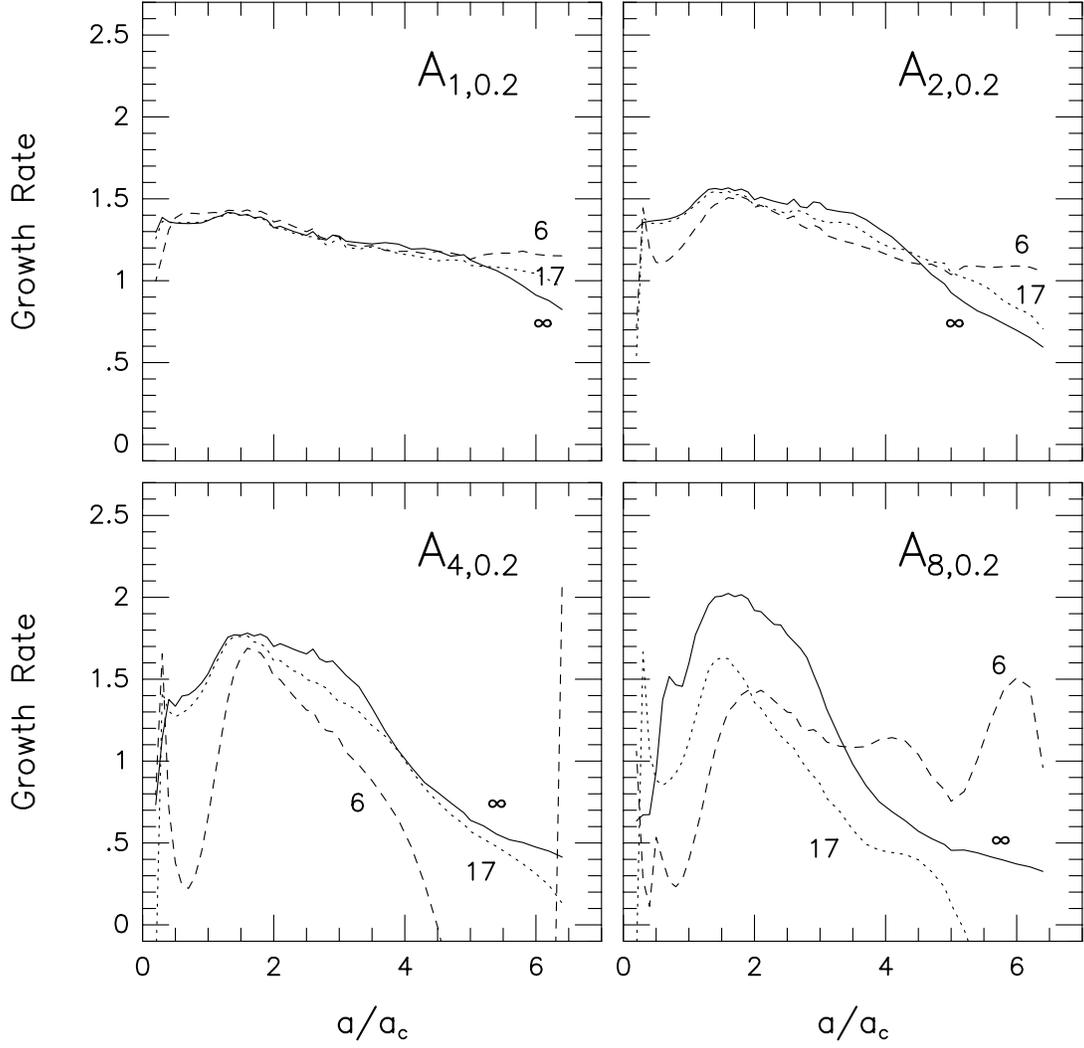

Fig. 16.— Dimensionless growth rate $\tilde{\Gamma}$ at $k = 2k_a$ vs. $a/a_c$, for antisymmetric modes with finite temperature, for $M = \infty$ (i.e. no thermal velocity) (solid curve), $M \approx 17$ (dotted), and $M \approx 6$ (short-dashed), for modes $A_{1,0.2}$, $A_{2,0.2}$, $A_{4,0.2}$, and $A_{8,0.2}$, as labelled.

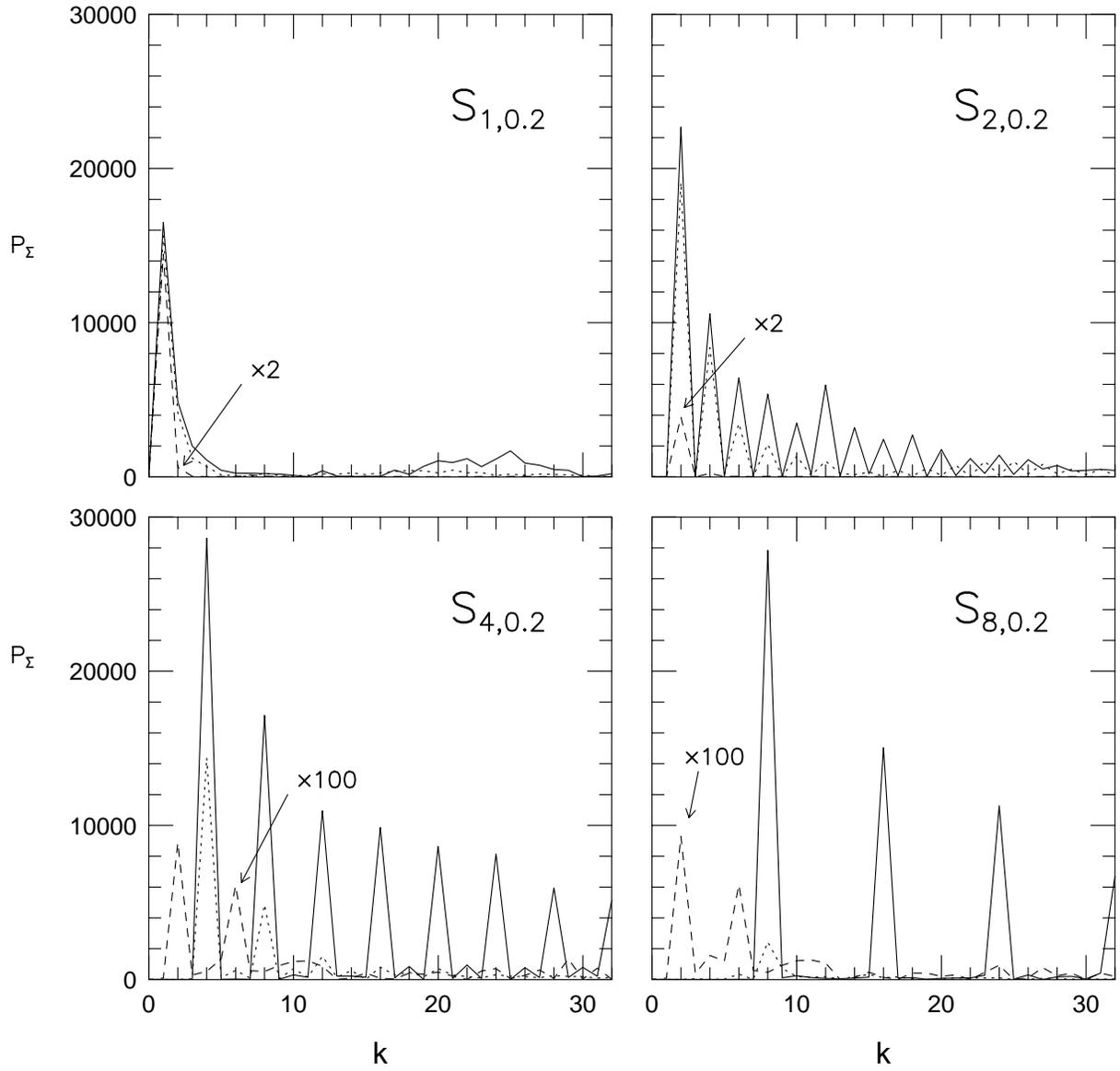

Fig. 17.— Same as Fig. 7, except for $\epsilon_s = 0.2$, and we display results in each panel for $M = \infty$(solid curve), $M \approx 17$(dotted), and $M \approx 6$(short-dashed).

One other detail must be noted in order to explain fully these results. Our numerical approach to representing finite temperature by dividing the mass at each initial location into several mass points with a randomly-selected thermal spread in velocities introduces a numerical source of additional perturbation, of small amplitude. This results from the discreteness of the mass distribution when a finite number of particles is used. The numerical noise caused by this effect introduces a spectrum of small-amplitude transverse perturbations with a broad range of wavenumber $k \geq 1$. The effect of this discreteness-generated perturbation spectrum is to add to the appearance of instability in the pancake, including a range of clumping scales evident at late times. Hence, there is significant power $P_\Sigma(k)$ at wavenumbers $k \neq k_s$ in the curves in Figure 17 for $a/a_c = 2$, even though the instability is still in the linear growth phase. This does not alter the description or the conclusions stated above regarding the fact that pancakes are unstable for $k > k_{\max}$ in the presence of finite temperature.

### 4.5. Filament Structure

Our simulations so far have shown that gravitational instability of pancakes subject to perturbation by either symmetric or antisymmetric modes can result in the production of dense bound clumps (or filaments in 2D). The degree of nonlinearity of the structures formed and their overdensity compared to the background at a given $a/a_c$ is a function of initial perturbation amplitude and wavenumber and the type of perturbation as discussed in the previous section. In this section we give a few statistics.

Our simulations show that a small, symmetric, transverse perturbation added to the fundamental pancake mode can result in filamentation within the pancake. For example, a symmetric transverse perturbation of wavelength equal to that of the fundamental pancake but with an amplitude equal to 10% of that of the fundamental pancake mode (i.e. mode $S_{1,0.1}$) can result in the production of one dense clump every $\lambda_p$ that contains 54% of the mass in the simulation box, with a maximum overdensity $(\delta\rho/\bar{\rho})_{\max} \sim 2000$ at $a/a_c = 7$. The $S_{1,0.2}$ mode, however, produces one clump every $\lambda_p$, but these contain 70% of the mass in the simulation box, with $(\delta\rho/\bar{\rho})_{\max} \sim 3000$, at the same epoch.

We have also investigated the density variation within the filaments formed in our simulations, as a function of the distance $r$ from the density maximum. We assumed a density profile of the form $\rho \propto r^{-m}$ and solved for $m$ by fitting $\log(\rho/\rho_{\mathrm{mean}})$ vs. $\log r$ to a straight line, where $\rho$ is the azimuthally-averaged value at a given $r$. Although this calculation is not extremely accurate, since it depends on the scheme by which the density is calculated by assignment of the mass of the simulation particles to the cells of a uniform grid (we use the CIC scheme) and also is a function of resolution, it will still give us a rough idea of the shape of the profile. In Figure 18, we have plotted the density profiles of the filaments which arose as a result of $S_{1,0,1}$ (open circles) and $S_{1,0,2}$ (filled circles) perturbations at $a/a_c = 7$. The fit to each data set is shown by the dotted and solid lines, respectively. We find $m \sim 1.1$ for the $S_{1,0,1}$ mode and $m \sim 1.2$ for $S_{1,0,2}$. Similarly the antisymmetric results for modes $A_{1,0,1}$ and $A_{1,0,2}$ are plotted in Figure 19. We find $m \sim 1.2$ for the $A_{1,0,1}$ mode and $m \sim 1.1$ for $A_{1,0,2}$, at $a/a_c = 7$. We must note, however, that in the antisymmetric mode, because of the ellipsoidal shape of the filament cross section, the results are not as accurate, since $\rho$ is an azimuthally-averaged value for each $r$.

We note that our results are in relatively good agreement with the analytical predictions of Fillmore & Goldreich (1985) who found a density profile $\rho \propto r^{-1}$ for a self-similar, purely cylindrical collapse. However, they are in even better agreement with the numerical results of Alimi et al. (1990), who found $m = 1.1 \pm 0.05$ for filaments formed at the intersection of two orthogonal pancakes of equal amplitude. The latter authors argued that the difference between their results and the predictions of Fillmore & Goldreich is real and originates from the fact that the filaments studied by Fillmore & Goldreich are uncompensated (i.e. overdensities not surrounded by compensating underdensities), whereas filaments formed at the intersections of pancakes are compensated. Since the perturbations we study in this paper are also compensated, we conclude that the small difference between our results and the predictions of Fillmore & Goldreich may also be real. Our results generalize the result of Alimi et al. (1990) to a much wider range of situations than the one they considered. Our filaments arise when pancakes are linearly perturbed, rather than as the effect of the nonlinear mode coupling of two equal amplitude pancakes studied by Alimi et al. As such, our results indicate that filaments evolve density profiles $\rho \sim r^{-m}$, $m = 1.1 \pm 0.1$ under a wider range of circumstances than either Fillmore & Goldreich or Alimi et al. considered.

## 5.  SUMMARY AND CONCLUSIONS

We have studied the gravitational instability of collisionless, cosmological pancakes through a set of idealized numerical N-body simulations by the Particle-Mesh method in 2D. We have investigated the response of 1D plane-wave pancakes to perturbations by either symmetric (density) or antisymmetric (bending or rippling) transverse modes, and compared our results to those of a thin-sheet energy argument analysis.

Our results can be summarized as follows: (1) Numerical N-body simulations reveal that cosmological pancakes are gravitationally unstable to transverse perturbation modes of wavenumber $k \gtrsim 1$ (i.e. wavelengths $\lambda \lesssim \lambda_p$), starting at epochs shortly after the formation of the first pancake caustics. This is a *linear* instability of the *nonlinear* pancake. (2) For the symmetric mode, power spectrum analysis of the pancake surface density perturbation indicates that for a given unstable $k_s$, the wavenumber of maximum power is $k = k_s$, while the maximum power occurs for $k = 2k_a$ for the antisymmetric mode. (3) During the linear phase of the instability, the dimensionless growth rate rises significantly above the value of 2/3 for such perturbations in the absence of the pancake mode and levels off prior to declining due to nonlinear saturation. The value at which it levels off converges for small initial amplitudes to a maximum value which is initial-amplitude-independent. This amplitude-independent value is the true linear growth rate, and its time-independent value prior to nonlinear saturation indicates that the linear instability is growing as a power-law in time, $\delta_\Sigma \propto t^\alpha$. The value of $\alpha$, and, hence, the linear growth rate, scales as $k^n$ where $n \approx 0.2-0.25$ for both the symmetric and antisymmetric modes (i.e. $\alpha_s \simeq \alpha_{s,1} k_s^n$ with $\alpha_{s,1} = 1.08$ and $n = 0.24$, while $\alpha_a \simeq \alpha_{a,1} k_a^n$ with $\alpha_{a,1} = 1.35$ and $n = 0.23$). (4) For equal values of initial perturbation amplitude, $\epsilon_a = \epsilon_s$, the antisymmetric mode has a more extended linear growth regime compared to the symmetric mode and, therefore, reaches the epoch of nonlinearity later. This is because the spatial maximum of the initial density perturbation at precaustic epochs is of first order in $\epsilon$ for the symmetric case, while it is only of second order for the antisymmetric case. (5) A comparison of these simulation results with the predictions of the thin-sheet energy argument indicates that the latter is incorrect in several significant respects. First, the energy argument predicts that instability will occur only for perturbation wavenumbers in the range $k_{\min} \leq k \leq k_{\max}$, where $k_{\min} \sim 1$ is the value below which Hubble expansion stabilizes the pancake, while $k_{\max}$ is the value above which the velocity dispersion of matter within the pancake central layer of caustics and shell-crossing prevents instability. Our simulation results are in good agreement with the prediction on the small wavenumber side (albeit, with a lower boundary for unstable values of $k$ which is not so sharp). However, we find no such agreement on the high wavenumber side. In fact, we find that velocity dispersion does *not* stabilize the pancake against perturbation by high wavenumbers $k > k_{\max}$. This is true for pancakes in an initially "cold" collisionless gas, in which the velocity dispersion is that from the purely 1D velocity along the direction of collapse, within the pancake central layer of counterstreaming matter. Such a result is analogous to that found previously by Hwang et al. (1989) for the instability of spherical shells which lead the expansion of a self-similar cosmological void in a cold collisionless gas, in which perturbations of all spherical eigennumbers $\ell$ were found to be unstable. However, here we have also considered the effects of perturbations on pancakes if the collisionless gas has a finite initial temperature. We find the remarkable result that, even the isotropic, random velocity dispersion of thermal motions is not sufficient to stabilize pancakes against perturbations of high wavenumber. In short, the prediction by the thin-sheet energy argument of stability for $k > k_{\max}$ is found numerically here to be invalid. For the unstable wavenumbers predicted by the thin-sheet energy argument, that argument moreover predicts a dimensionless growth rate which scales with perturbation wavenumber as $k^{1/2}$. Our simulation results find, instead, a growth rate scaling as $k^n$ where $n \approx 0.2 - 0.25$ (7) The nonlinear outcome of the instability found here is the prediction of strongly nonlinear bound clumps (which are filaments in our 2D simulations) in the pancake mid-plane, one per $\lambda_s$ for symmetric modes and two per $\lambda_a$ for antisymmetric modes. (5) The "clumps" or filaments resulting from symmetric modes have a circularly symmetric cross section, while those resulting from antisymmetric modes have an elongated and ellipsoidal cross section. (8) The azimuthally-averaged density profile of these filaments is a power-law in distance from the filament center, $\rho \sim r^{-m}$, $m = 1.1 \pm 0.1$.

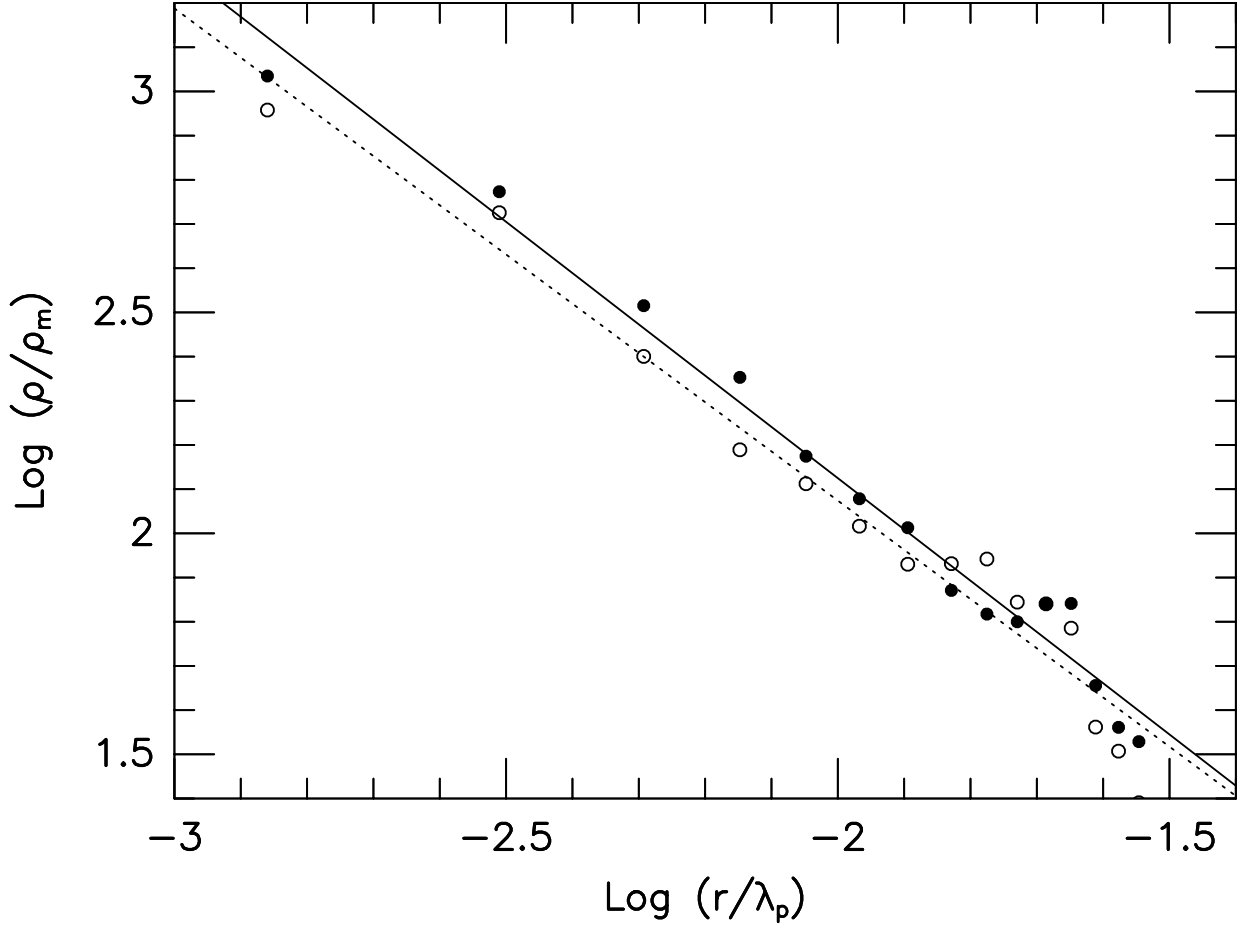

Fig. 18.— Azimuthally-averaged density profile of the filaments produced by mode $S_{1,0,1}$ (open circles) and $S_{1,0,2}$ (filled circles) at $a/a_c = 7$. We plot $\log(\rho/\rho_m)$, where $\rho_m$ is the mean density, versus $\log(r/\lambda_p)$, where $r$ is the distance measured from density peaks at the filament center. The dotted and solid lines are the fit to the $S_{1,0,1}$ and $S_{1,0,2}$ modes, respectively.

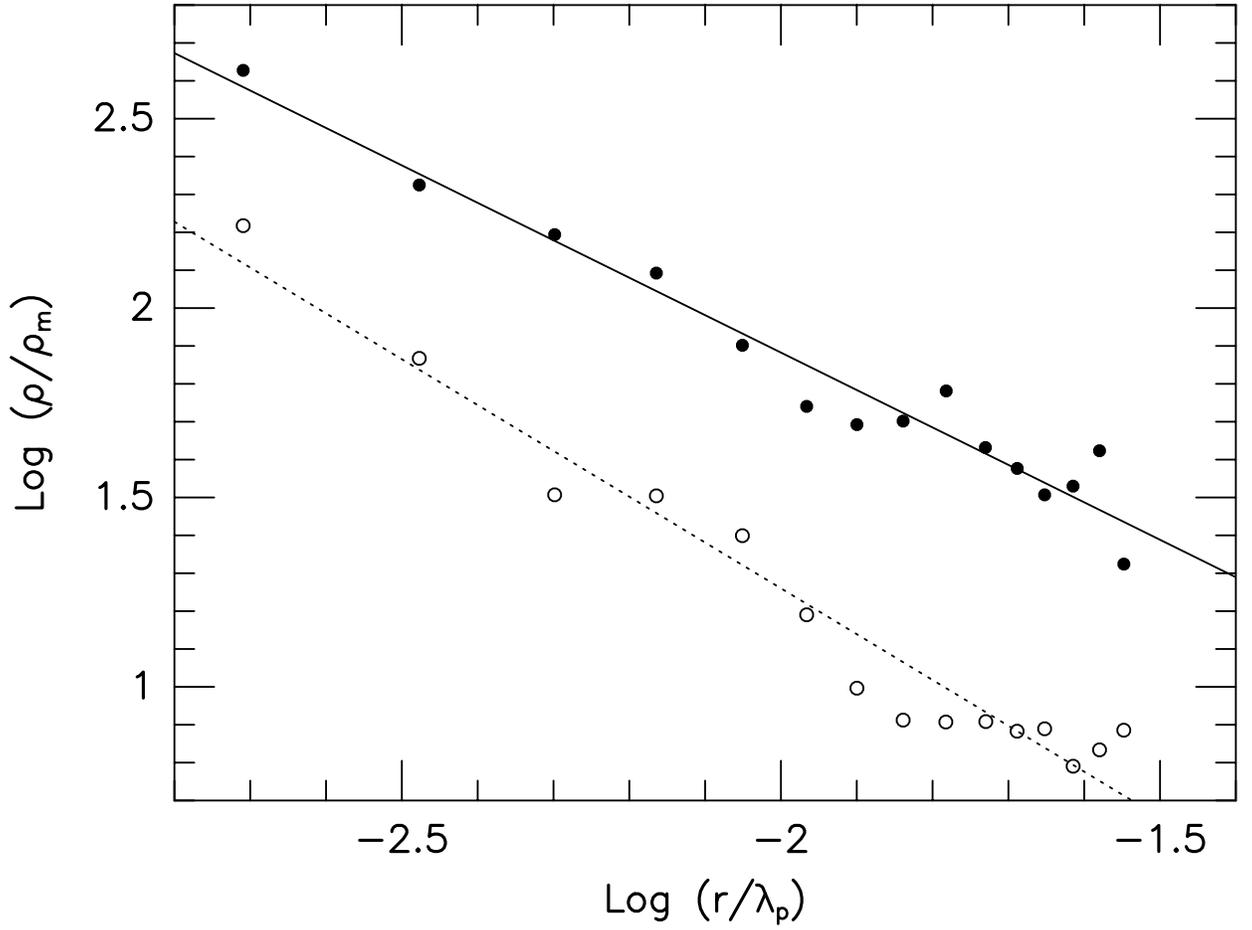

Fig. 19.— Same as Fig. 18, except for modes $A_{1,0.1}$ and $A_{1,0.2}$, respectively.

We are especially grateful to Jens Villumsen for providing the 2D Particle Mesh code (as a collaborator in the development of our ASPH/PM code) which we modified for use in this paper. We thank M. Umemura and P. J. E. Peebles for discussions. A.V. would like to thank NSF and the Ministry of Education, Science, and Culture of Japan for a joint summer graduate fellowship in 1994 while part of this work was in progress. P.R.S. was a participant in the NSF supported Workshop on Galaxy Formation at The Institute for Theoretical Physics, UCSB, Santa Barbara, in 1995 during part of this work. The calculations presented in this paper were performed at the High Performance Computing Facility at The University of Texas. We are pleased to acknowledge the support of NASA Grants NAGW-2399 and NAG5-2785, Robert A. Welch Foundation Grant F-1115, and Cray Research.